\documentclass[pre,floatfix, superscriptaddress,a4paper,nofootinbib]{revtex4}
\usepackage{amsmath,bm,graphicx}
\usepackage{amssymb}
\usepackage{amsfonts}
\usepackage{epstopdf}
\usepackage{mathrsfs}
\usepackage{color}
\usepackage{latexsym}
\usepackage{hyperref}
\usepackage{subfigure}
\usepackage{bm}
\usepackage{natbib}

\begin{document}

%\title{Simple formula for the transport across energetic and entropic barriers in the linear response regime}

\title{Closed formula for the transport of micro- nano-particle across model porous media}

\author{Paolo Malgaretti}
\email[Corresponding Author: ]{p.malgaretti@fz-juelich.de }
\affiliation{Helmholtz Institute Erlangen-N\"urnberg for Renewable Energy (IEK-11), Forschungszentrum J\"ulich, Erlangen, Germany}

\author{Jens Harting}
\affiliation{Helmholtz Institute Erlangen-N\"urnberg for Renewable Energy (IEK-11), Forschungszentrum J\"ulich, Erlangen, Germany}
\affiliation{Department of Applied Physics, Eindhoven University of Technology, Eindhoven,
The Netherlands
}
\begin{abstract}
In the last decade the Fick-Jacobs approximation has been exploited
to capture the transport across constrictions. Here, we review the derivation
of the Fick-Jacobs equation with particular emphasis on its linear response
regime. We show that for fore-aft symmetric channels the flux of
non-interacting systems is fully captured by its linear response regime. For
this case we derive a very simple formula that captures the correct trends and
that can be exploited as a simple tool to design experiments or simulations.
Finally, we show that higher order corrections in the flux may appear for
non-symmetric channels.
\end{abstract}
\maketitle

\section{Introduction}

It is common to experience long queues form when a constriction occurs on a highway~\cite{Lighthill1955,Wang_traffic}. 
Such an (unlucky) phenomenon is clearly the result of the ``local'' confinement: due to the constriction, vehicles slow down hence reducing the local ``mass'' flux as compared to the clear part of the highway. Such a local reduction of the mass flow causes the onset of the annoying queues that every now and then we experience.
This phenomenon does not occur only on highways. It becomes a major issue close to emergency exits in the case of panic~\cite{Vermuyten_review}.
The very same dynamics occurs also at smaller scales and for simpler systems. For example, it is common experience that it is difficult to extract pills from a container if the opening is too small. Here, pills tend to ``clog'' i.e., to form stable structures close to the opening of the container that prevent pills from going out. 
The very similar dynamics occurs in silos containing crops~\cite{Jeong2018}, in erosion~\cite{Jaeger2018}, in suspensions of hard and soft particles~\cite{Marin2018,KVKHS14,BAHK21}, in herds of sheep~\cite{Zuriguel2015}, 
in the onset of panic in ants~\cite{Altshuler2005}, and even humans~\cite{Zuriguel2020}.

The effect of confinement does not have to be unpleasant, as it is for traffic jams, or inconvenient, as it is for clogging of silos. 
Vice versa, tuning the shape of the confining media can be an intriguing and novel way to control the dynamics of the confined system. 
%exploited to control the dynamics of the confined system. 
For example, microfluidic devices exploit variations of the section of the micro channels they are made of to control the dynamics of fluid and to induce the formation of droplets~\cite{Squires2005,Dressaire2017,Doufene2019,Convery2019}.  
Similarly, Tunable Resistive Pulse Sensing (TRPS) techniques exploit micro- nano-pores to analyze small particles ranging from a few tens of nanometers up to micrometric scale~\cite{Willmott2015}. In particular, TRPS has been used to direct detect antibody-antigen binding~\cite{Saleh2003}, to measure elecrophoretic mobility of colloidal particles~\cite{Ito2004}, to perform single-molecule detection~\cite{Heins2005} and to measure the zeta-potential of nanometric particles~\cite{Arjmandi2012}.
Alternatively, Chromatography techniques have been developed to separate micro- or nano-particles depending on both their size as well as their surface properties~\cite{RobardsBook,Reithinger2011,Michaud2021,SMKK2008}.
Finally, at even smaller scales, nanopores have been designed to sequence DNA molecules~\cite{Gautam2010}.

Transport in confinement is not relevant only for particle detection/analysis.
Indeed, the flow of fluids across a porous medium is crucial in diverse
scenarios. For example, oil recovery industries have put much effort into
developing techniques to maximize the extraction of oil from the rock matrix it
is embedded in~\cite{Carvalho2015,Foroozesh2020}. Similarly, understanding the
dependence of the flow of water on the porosity of the soil is crucial in
environmental sciences~\cite{Farhadian2019}. Moreover, diverse techonlogies related to the energy transition such as blue-energy~\cite{Roij2011}, hydrogen technology~\cite{Preuster2017,Solimosi2022}, electrolyzers and fuel cells~\cite{Suter2021,Du2022}, or $CO_2$ segregation~\cite{Hepburn2019} rely on the transport of (charged) chemical species across nanoporous materials.

Finally, several biological systems are controlled by the transport of confined
complex fluids. For example, neuronal transmission relies on the transport of
neuro-receptors among neurons and to their specific binding
sites~\cite{Albers_book}. Moreover, cell regulation relies on the proper tuning
of the concentrations of electrolytes inside the cell. Such a regulation occurs
via dedicated pores and channels whose shape makes them very sensitive to
specific ions~\cite{Pethig1986,Dubyak2004,Calero,Roth2014,Peukert2017} and RNA
is transported across the nuclear
membrane~\cite{Gracheva2017,Bacchin2018,Dagdug2019}.  Moreover, the lymphatic
and circulatory systems in mammals rely on the transport of quite heterogeneous
suspensions composed of a variety of components, spanning from the nanometric
size of ions up to the micrometric  size of red blood cells, across
varying-section elastic
pipes~\cite{KVKHS14,Nipper2011,Wiig2012,Yoganathan1344}. Finally, the survival
of plants relies, at large scales, on the proper circulation of liquid (sap)
along the trunk~\cite{RMP_Plants2016} and at short scales on the cytoplasmic
streaming within the cells~\cite{Shimmen2004}.

All the above mentioned systems rely or depend on the dynamics under confinement.
Therefore, understanding the dynamics and transport properties of confined
complex systems such as ions, molecules, polymers, colloidal particles, and
suspensions is of primary importance for the understanding of a wide spectrum
of phenomena and for the development of technological applications.  Even more,
identifying the relevant parameters controlling key features, like transport or
phase transitions, will open a new route for controlling the dynamics of
confined systems upon tuning the geometry of the confining media. 

Up to now, there has been no systematic study of the dependence of the dynamics
of confined systems upon changing the shape of the confining walls.  The main
reason is the large effort that such a study requires.
Indeed, \textit{experimentally} tuning the shape of a pore is a tremendous task since, if possible at all, it requires to synthesize every time a new item from scratch.
On the \textit{theoretical} side, studying the dynamics and the transport of
confined systems is a tremendous task since it requires to capture several
length, time and energy scales. In fact, the length scales range from the
nanometric scale, typical for ions and for van der Waals interactions to the
micrometric scale of colloids, polymers and macromolecules up to the
millimeters/centimeters scale of microfluidic devices. Concerning time scales,
the spectrum spans the diffusion time of small particles and ions over their
size $\sim \mu \text{sec}$ up to the long time scales typical of transport
$\sim \text{sec}$. Concerning energy scales, they range from thermal energy
$k_BT$ ($\sim 10^{-21} \text{J}$) up to van der Waals and electrostatic
interactions whose magnitude can be of several $k_BT$. On the top of these
``direct'' interactions also the effective interactions induced by the
confinement should be accounted for. For example, squeezing a deformable
object, like a polymer or a vesicle, through a constriction can require
quite an amount of energy that can easily reach the order of $100-1000\, k_BT$.
Given such a complexity, one typically would rely on numerical techniques such
as molecular dynamics. However, the wide range of interactions (Van der Walls,
electrostatic..) jointly with the wide range of time and length scales imposes
to put forward numerical approaches capable of properly resolving the smallest
length, time and energy scales.  At the same time, such an approach should also
resolve the large length, time and energy scales.  Accordingly, the numerical
route becomes quite demanding from the perspective of the computational time.

Since both experimental and numerical routes are quite expensive, an
approximated analytical route based on some controllable expansions may become
appealing. 
Intriguingly, it is possible to obtain simple analytical models that capture
some features of the dynamics of confined systems. The key idea,  is to
``project'' the dynamics of the system onto some relevant coordinate (in
chemistry sometimes called ``reaction coordinate'') and then to study the
dynamics of these few (typically one) degrees of freedom.  For example, in the
case of polymer translocation across pores, the most important observable is
the time the polymer takes to cross from one side to the other of the pore.
Therefore, the relevant degree of freedom is the position of the center of mass
of the polymer whereas the degrees of freedom associated with the position of
the monomers can be integrated out. 

In this contribution, we briefly review the derivation of the Fick-Jacobs
approximation~\cite{Zwanzig,Reguera2001,Kalinay2005,Kalinay2005_2,Kalinay2008,Martens2011,Dagdug2013,Malgaretti2013}
and its use in studying transport across corrugated pores and channels.
The Fick-Jacobs approximation has been shown to be applicable to the transport
of
ions~\cite{Malgaretti2014,Malgaretti_macromolecules,Malgaretti2015,Chinappi2018,Malgaretti2019_JCP},
colloids~\cite{Reguera2006,Reguera2012,Marconi2015,Malgaretti2016_entropy,Puertas2018},
rods~\cite{Malgaretti2021},
polymers~\cite{Bianco2016,Malgaretti2019,Ceccarelli2019}, and more recently
even active systems~\cite{Malgaretti2017,Kalinay2022,Antunes2022}, chemial
reactors~\cite{Santamaria2016} and pattern-forming systems~\cite{Chacon2020}. 
In the following we re-derive the Fick-Jacobs approximation with particular
emphasis on the regime in which the current is proportional to the applied
force. In such a regime, it is possible to derive a closed formula that
accounts for the dependence of the flux on the geometry of the channel.
Interestingly, our derivation naturally highlights a few relations between the
underling Smoluchowski equation and the linear response theory. Even though
this work is motivated by the transport in confined pores and channels, the
results we derive are valid for all $1D$ systems (independently of the physical
origin of the effective potential) in the dilute regime (for which mutual
interactions can be neglected) and whose dynamicsis governed by the Smoluchowski
equation (i.e.~in the overdamped regime). 

\begin{figure}
\includegraphics[scale=0.5]{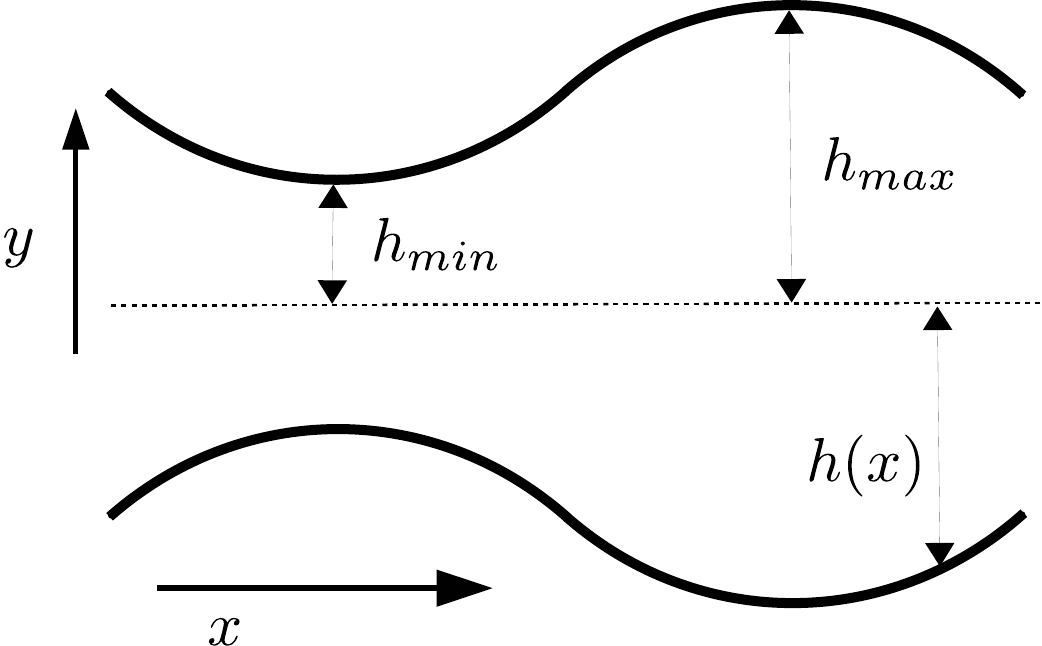}
\caption{Cartoon of the varying-section channel.}
\label{fig:cartoon_analytical}
\end{figure}

%show, in a simple way, that the flux of the non-interacting system is linear in the external force. We remark that such linearity is granted since the starting equation is linear in the force. However we find instructive to show it explicitly.

\section{Model}
In the following we are interested in the transport of a single colloidal particle confined in an axially symmetric channel characterized by its half section (see Fig.~\ref{fig:cartoon_analytical} for a scketch of the system)
\begin{align}
h(x)=h_0 +h_1\cos\left(2\pi \frac{x}{L}\right)\,.
\end{align}
The time evolution of the probability density is governed by the Smoluchowski equation
\begin{align}
\dot{\rho}(\mathbf{r},t)=\nabla\cdot \left[D\nabla\rho(\mathbf{r},t)+D\beta \rho(\mathbf{r},t) \nabla W(\mathbf{r})  \right]\,,
\label{eq:adv-diff}
\end{align}
where $D$ is the diffusion coefficient, $\beta^{-1}=k_BT$ is the inverse thermal energy, $k_B$ the Boltzmann constant, $T$ the absolute temperature and 
\begin{align}
W(\mathbf{r})=\begin{cases}
\phi(\mathbf{r}) & |r|< h(x)\\
\infty   & \text{else}
\end{cases}
\end{align}
is the effective potential responsible for both confining the particle within the channel and for additional soft interactions, $\phi(\mathbf{r})$ with the channel walls.
For smoothly-varying channel cross-sections, $\partial_x h(x)\ll 1$, it is possible to factorize the probability density~\cite{Zwanzig,Reguera2001,Kalinay2008,Martens2011,Dagdug2013,Malgaretti2013}
\begin{align}
\rho(\mathbf{r},t)=p(x,t)\dfrac{e^{-\beta W(\mathbf{r})}}{e^{-\beta A(x)}}\,,
\end{align}
where 
\begin{align}
A(x)=-k_BT\ln\left[\frac{1}{\pi h_0^2}\int_{-\infty}^{\infty} e^{-\beta W(\mathbf{r})}r dr\right]
\end{align}
is the local free energy~\cite{Malgaretti2016}. 
Moreover, integrating along the radial direction leads to
\begin{align}
\dot{p}(x,t)=\partial_x\left[D\partial_x p(x,t)+D\beta p(x,t)\partial_x A(x) \right]\,.
\end{align}
Such a procedure is called {\it Fick-Jacobs
approximation}~\cite{Zwanzig,Reguera2001,Malgaretti2013}. Its regime of
validity has been assessed by several
groups~\cite{Reguera2006,berezhkovskii2007diffusion,Burada2007,dagdug2015,Kalinay2005,Kalinay2005_2,Kalinay2006,Martens2011,Dagdug2012_2,Dagdug2015_2}.
In particular, it has been shown that the quantitative reliability of the
Fick-Jacobs approximation can be enhanced by introducing a position dependent
diffusion
coefficient~\cite{Reguera2006,berezhkovskii2007diffusion,Burada2007,dagdug2015,Kalinay2005,Kalinay2005_2,Kalinay2006,Martens2011,Dagdug2012_2,Dagdug2015_2},
$D(x)$, hence leading to the set of equations
\begin{align}
\dot{p}(x,t)&=-\partial_x J(x,t)\\
\frac{J}{D(x)}&=-\partial_x p(x)-\beta p(x) \partial_x A(x)\,.
\label{eq:def-flux}
\end{align}
%where $D(x)$ is the diffusion coefficient that, for generality, we assume that it may be position dependent and $A(x)$ is an effective potential that accounts for all conservative forces acting on the system.
Eq.~\eqref{eq:def-flux} is completed with the  boundary conditions
\begin{align}
p(-L) &=p(L)\label{eq:P-BC}\\
\int_{-L}^{L} p(x)dx &=1. \label{eq:app-norm}
\end{align}
We decompose the effective force $-\partial_x A(x)$ as the net force
\begin{align}
f=-\frac{1}{2L}\int_{-L}^L \partial_x A(x) dx=-\frac{\Delta A}{2L}
\end{align}
and 
\begin{align}
A_{eq}(x)=A(x)+fx.
\end{align}
%where 
%\begin{align}
%A(x)=A_{eq}(x)-fx
%\end{align}
$f$ accounts for the net force responsible of the flux and $A_{eq}(x)$ accounts for all the other conservative forces that will not give rise to any flux. %(see Fig.\ref{fig:scheme_pot}). %, encoded in the potentia $A_{eq}$, as well as for external forces, $f$. 
In the following, we expand both the flux, $J$, and the density, $p$, about the equilibrium case:
\begin{align}
J=&J_0+J_1+J_2+...\\
p(x)=&p_0(x)+p_1(x)+p_2(x)+...
\end{align}
Note that due to Eq.~\eqref{eq:app-norm} at zeroth order we have
\begin{align}
\int_{-L}^{L} p_0(x)dx&=1\,. \label{eq:app_norm-p0}
\end{align}
This implies
\begin{align}
\int_{-L}^{L} p_n(x)dx&=0\,\,\,\,\forall n\neq 0
\label{eq:app_norm-pn}
\end{align}

Accordingly, at order zero we have
\begin{align}
p_0(x)&=\tilde p e^{-\beta A_{eq}(x)}\label{eq:app_p0}\\
J_0&=0\label{eq:J0}\\
\tilde p &= \frac{1}{\int_{-L}^{L}e^{-\beta A_{eq}(x)}dx}\,.\label{eq:p-tilde}
\end{align}
At the generic $n$-th order we have
\begin{align}
\frac{J_n}{D(x)}=-\partial_x p_n(x)-\beta p_n(x) \partial_x A_{eq}(x)+\beta p_{n-1}(x)f\,,
\end{align}
the solution of which reads
\begin{align}
p_n(x)=e^{-\beta A_{eq}(x)}\left[\int\limits_{-L}^x \left[\beta p_{n-1}(y)f-\frac{J_n}{D(y)}\right]e^{\beta A_{eq}(y)}dy+\Pi_n \right]\,.
\label{eq:pn}
\end{align}
Here, $J_n$ and $\Pi_n$ are integration constants.
Imposing the periodic boundary conditions, $p_n(-L)=p_n(L)$, and recalling that $A_{eq}(-L)=A_{eq}(L)$ leads to
\begin{align}
\int_{-L}^{L} \left(\frac{J_n}{D(y)}-\beta p_{n-1}(y)f\right)e^{\beta A_{eq}(y)}dy=0\,,
\end{align}
with
\begin{align}
J_n=\beta f \dfrac{\int_{-L}^{L}  p_{n-1}(y)e^{\beta A_{eq}(y)}dy}{\int_{-L}^{L} \dfrac{e^{\beta A_{eq}(y)}}{D(y)}dy}= \beta f \tilde{p} \dfrac{\int_{-L}^{L}  \frac{p_{n-1}(y)}{p_0(y)}dy}{\int_{-L}^{L} \dfrac{e^{\beta A_{eq}(y)}}{D(y)}dy}\, .
\label{eq:J_n}
\end{align}
In the last step we used Eq.~\eqref{eq:app_p0}. 
%\begin{figure}
%\includegraphics[scale=0.5]{scheme_pot.eps}
%\caption{Cartoon of a particle (orange circle) in a dashboard potential $A(x)$ (blue solid line). $\Delta A$ is the overall decay of the potential over the length $L$.}
%\label{fig:scheme_pot}
%\end{figure}
Finally, $\Pi_n$ is determined by imposing Eqs.~\eqref{eq:app_norm-p0},~\eqref{eq:app_norm-pn}
\begin{align}
\Pi_n=-\tilde{p}\int_{-L}^L e^{-\beta A_{eq}(x)}\int\limits_{-L}^x \left[\beta p_{n-1}(y)f-\frac{J_n}{D(y)}\right]e^{\beta A_{eq}(y)}dy dx\,.
\label{eq:Pi}
\end{align}
At leading order in the force, Eqs.~\eqref{eq:pn}~\eqref{eq:J_n} read
\begin{align}
p_1(x)=&\, e^{-\beta A_{eq}(x)}\left[\beta f \tilde{p}(x+L) -J_1 \int_{-L}^x \frac{e^{\beta A_{eq}(y)}}{D(y)}dy\right] \label{eq:p1}\,,\\
J_1=&\, \dfrac{2\beta f L}{\int_{-L}^{L}e^{-\beta A_{eq}(x)}dx \int_{-L}^{L} \frac{e^{\beta A_{eq}(x)}}{D(x)}dx}\,. \label{eq:flux}
\end{align}
Interestingly, from Eq.~\eqref{eq:flux}, it is possible to identify a force-independent channel permeability 
\begin{align}
\chi = \dfrac{ 2 \beta  L}{\int_{-L}^{L}e^{-\beta A_{eq}(x)}dx \int_{-L}^{L} \frac{e^{\beta A_{eq}(x)}}{D(x)}dx}\,.
\label{eq:chi}
\end{align}
As expected, Eq.~\eqref{eq:chi} agrees with the derivation of the effective diffusion coefficient for a particle at equilibrium and in the presence of entropic barriers~\cite{Lifson1962,Rubi2001}. This is in agreement with the linear response theory within which the transport coefficients that determine the flux under external forces can be determined from equilibrium properties. 

Some general remarks can be derived in the case of fore-aft symmetric channels, for which $A_{eq}(x)=A_{eq}(-x)$, and diffusivities, $D(x)=D(-x)$. For such cases, the magnitude of the flux should depend solely on the magnitude of the force and not on its sign. This implies that 
\begin{align}
J_{2n}=0,\quad \forall n>0 \label{eq:J2n}\,.
\end{align}
In order to proceed, we recall that, for fore-aft symmetric $f(x)$ and $g(x)$, the following equality holds:
\begin{align}
\int_{-L}^{L} g(x) \int_{-L}^x f(y) dy dx = \frac{1}{2} \int_{-L}^L f(x) dx \int_{-L}^L g(x) dx\,
\label{eq:sum}
\end{align}
Enforcing the condition in Eq.~\eqref{eq:J2n} into Eq.~\eqref{eq:J_n} and using the last expression leads to 
\begin{align}
\Pi_{n}=0,\quad \forall n>0\,.
\end{align}
and, substituting again into Eq.~\eqref{eq:J_n} eventually leads to 
\begin{align}
J_{n}=0,\quad \forall n\geq 1 \label{eq:J2n1}\,.
\end{align}
Interestingly, we note that even though $\Pi_{n>0}=0$ and $J_{n>1}=0$ the density profile is still sensitive to higher order corrections in the force, i.e. in general $p_n\neq 0$.
According to this analysis, Eq.~\eqref{eq:flux} is not just the linear contributions to the flux rather it provides the exact expressions at every order in the external force. 
The outcome of this analysis is indeed intuitive since it states that for non-interacting systems confined within fore-aft symmetric channels non-linear effects are absent. The same results are indeed valid for any $1D$ problem with such a symmetry.

In contrast, if neither the potential, $A(x)$, nor the diffusion profile, $D(x)$, have a defined parity, then the left-right symmetry is broken, Eq.~\eqref{eq:J2n} does not hold anymore, and indeed a diode effect may set for sufficiently large external forces. 
We can assess the dependence of the diode effect on the geometry of the channel by calculating
\begin{align}
J_2 =\beta f \dfrac{\int\limits_{-L}^L \int\limits_{-L}^x \!\!\! \beta \tilde{p}f-\frac{J_1}{D(y)}e^{\beta A_{eq}(y)}dy+\Pi_1 dx}{\int_{-L}^{L} \dfrac{e^{\beta A_{eq}(y)}}{D(y)}dy}\, .
\end{align}
Using 
\begin{align}
\Gamma(x)=\int_{-L}^{x} \dfrac{e^{\beta A_{eq}(y)}}{D(y)}dy
\end{align}
and the definition of $J_1$ we obtain
\begin{align}
J_2 =\frac{\beta f}{\Gamma(L)} \int\limits_{-L}^L\beta \tilde{p}f(x+L)-2\beta \tilde{p}f L \dfrac{\Gamma(x)}{\Gamma(L)}+\Pi_1 dx\,.
\end{align}
Finally, using the definition of $\Pi_1$ we obtain 
\begin{align}
J_2 =& \frac{(\beta f L)^2 \tilde{p}}{\Gamma(L)} \frac{1}{L}\int\limits_{-L}^L \left[\left(\frac{x}{L}+1\right)- 2\dfrac{\Gamma(x)}{\Gamma(L)}\right]\left[1-e^{-\beta A_{eq}(x)}\right] dx\,.
\end{align}
\subsection{Transport across free energy barriers}
In the case of transport of point-like particles across $3D$ varying-section channels with axial symmetry the effective potential reads
\begin{align}
A^{(id)}_{eq}(x)=-2 k_BT \ln \left[\frac{h(x)}{h_0}\right]\, ,
\end{align}
where $h(x)$ is the local half-section of the channel and $h_0$ its average value (see Fig.\ref{fig:cartoon_analytical}). 
Accordingly, Eq.~\eqref{eq:flux} reads
\begin{align}
J_{id}=\dfrac{2\beta f L}{\int_{-L}^{L} \frac{h^2(x)}{h_0^2}dx\int_{-L}^{L} \frac{h_0^2}{h^2(x)D(x)}dx}\,.
\label{eq:flux_id}
\end{align}
%where we used
%\begin{align}
%\int_{-L}^{L}\frac{h(x)}{h_0}dx =L
%\end{align}
In the case of micro- or nano-particles that undergo solely excluded volume interactions with the channel walls, the effective channel half-section becomes $h(x)-R$ where $R$ is the particle size and we obtain
\begin{align}
A^{(pcl)}_{eq}(x)=-2 k_BT \ln \left[\frac{h(x)-R}{h_0}\right]\,,
\end{align}
which leads to
\begin{align}
J_{pcl}=\dfrac{2\beta f L}{\int_{-L}^{L} \frac{(h(x)-R)^2}{h_0^2}dx\int_{-L}^{L} \frac{h_0^2}{(h(x)-R)^2 D(x)}dx}\,.
\label{eq:flux_pcl}
\end{align}
%where we used
%\begin{align}
%\int_{-L}^{L}\frac{h(x)-R}{h_0}dx =L-RL/h_0
%\end{align}
We recall that $R<h_0-h_1$ for the particle to be able to cross the channel. 
Finally, several groups have shown that the Fick-Jacobs approximation can be improved by assuming a position-dependent diffusion coefficient~\cite{Zwanzig,Reguera2001,Kalinay2006,Kalinay2008,Martens2011,Dagdug2012_2,dagdug2015,Dagdug2015_2}. Nowadays, there is general agreement that  the approximated formula for the diffusion coefficient reads~\cite{Reguera2001} (or is in practice equivalent to)
\begin{align}
D(x)=\dfrac{D_0}{\sqrt{1+(\partial_x h(x))^2}}\,.
\label{eq:diff}
\end{align}

\begin{figure*}
\includegraphics[scale=0.47]{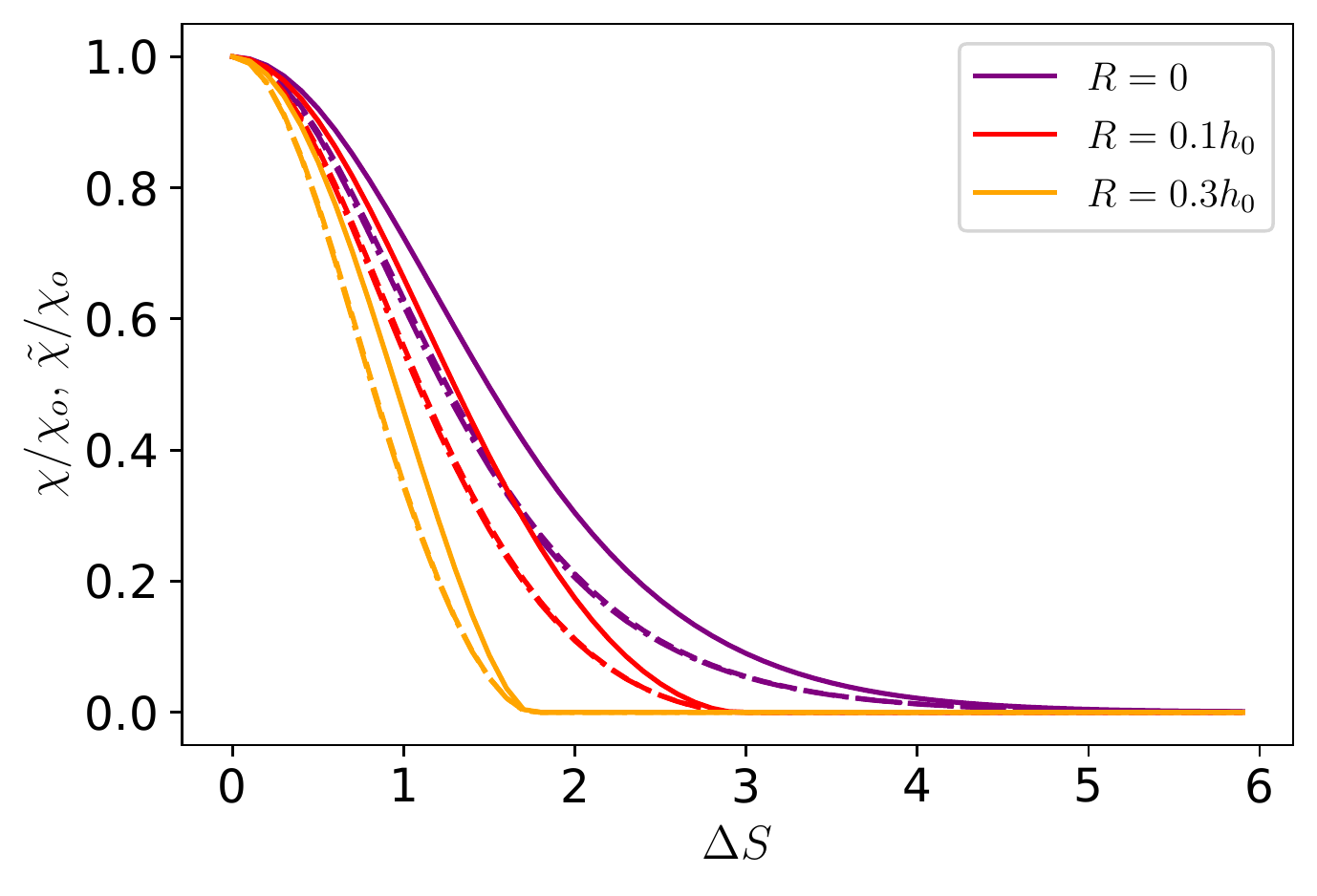}\quad\quad \includegraphics[scale=0.47]{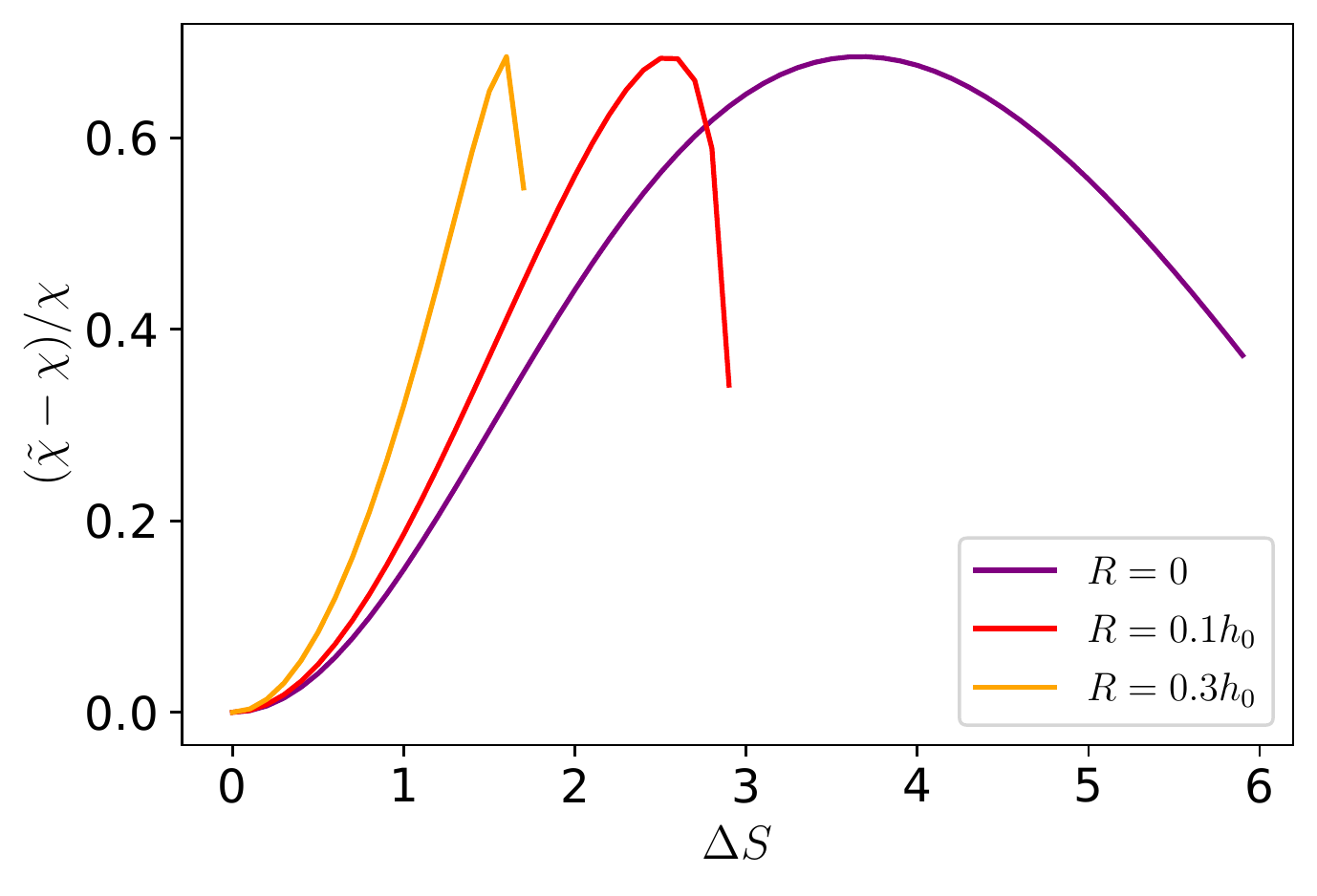}
\includegraphics[scale=0.47]{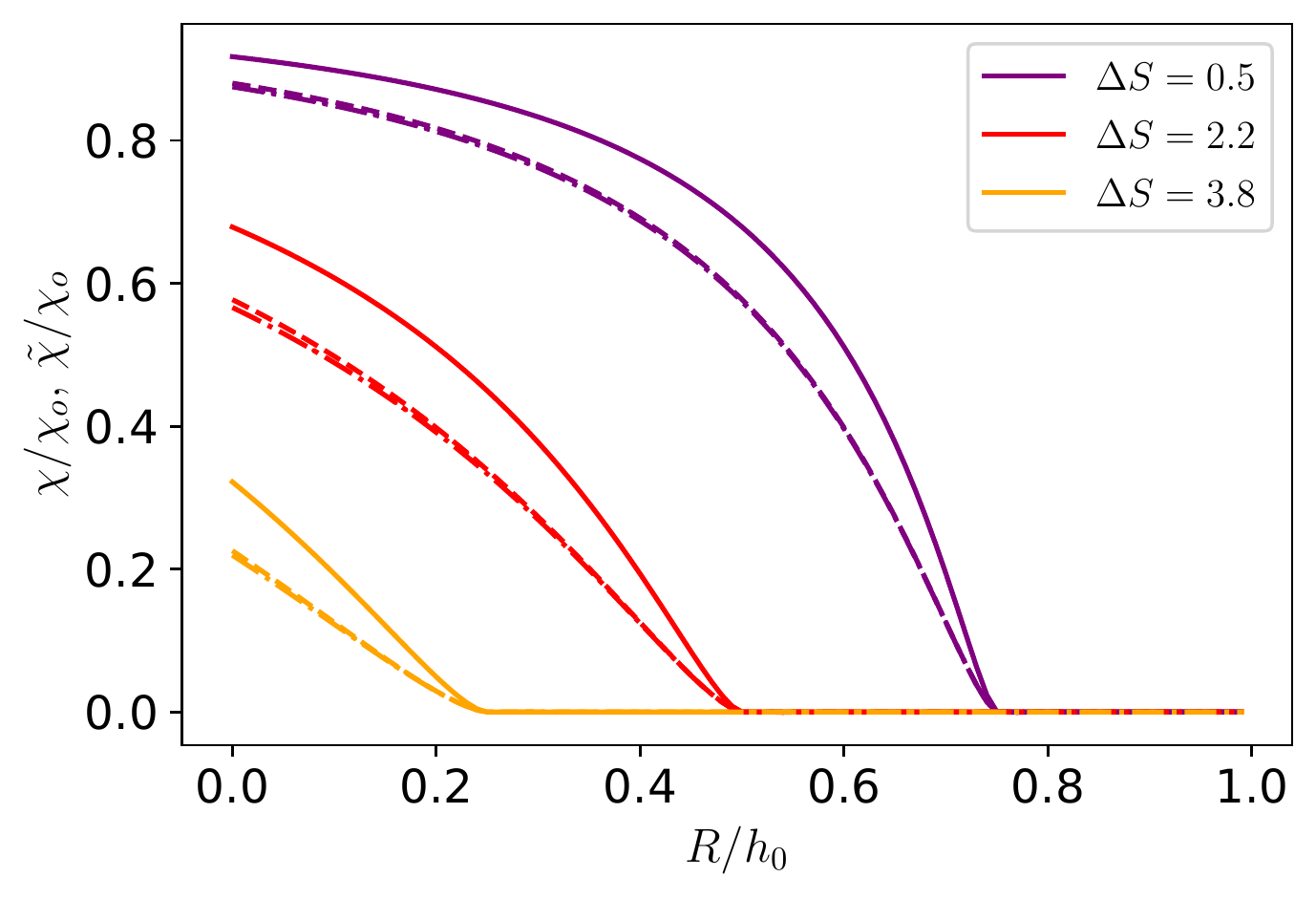}\quad\quad \includegraphics[scale=0.47]{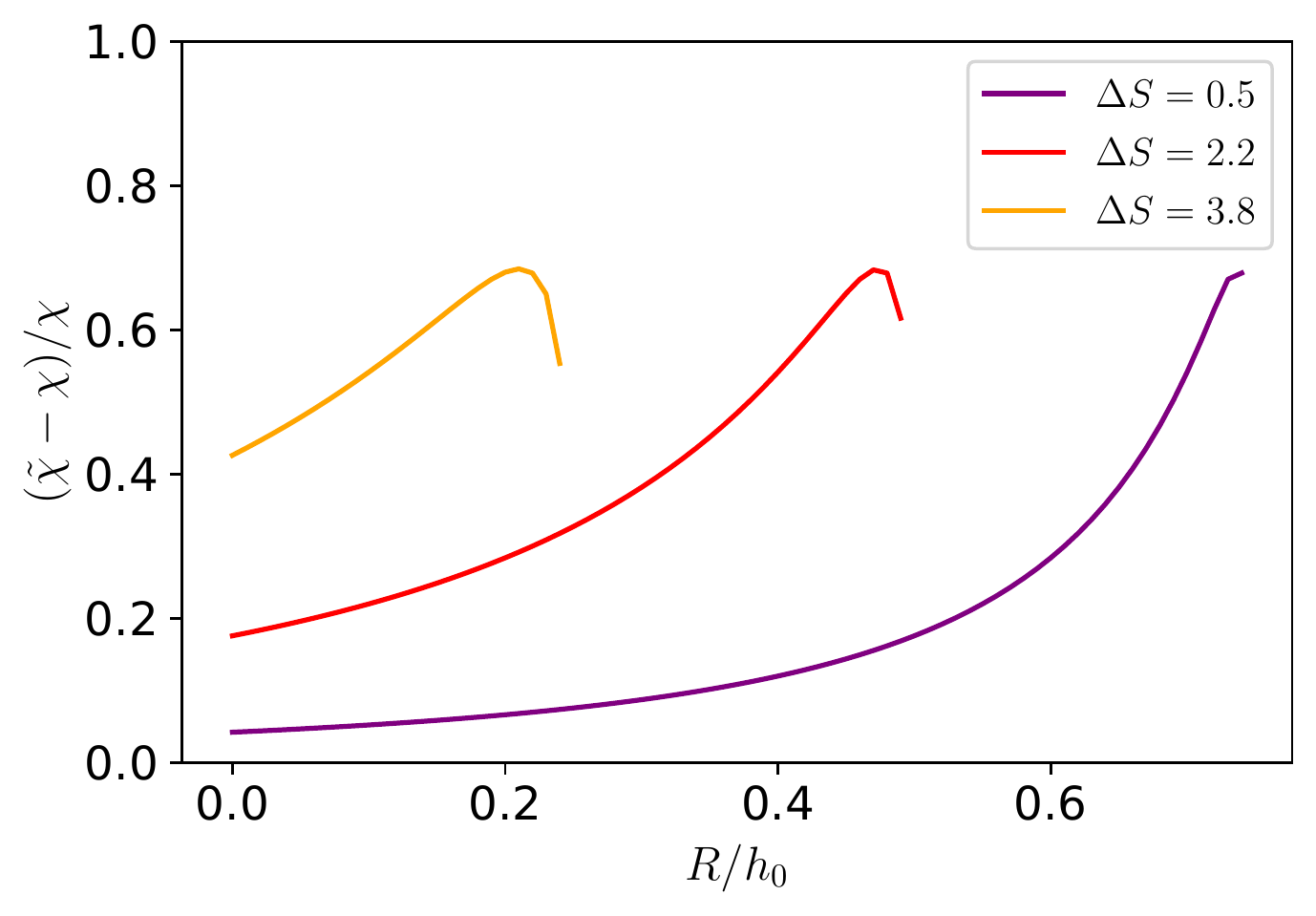}
\caption{Transport across porous media. Upper left: permeability, $\chi$, as obtained form Eq.~\eqref{eq:flux-simple} (solid lines), Eq.~\eqref{eq:flux} with constant diffusion coefficient (dashed lines) and Eq.~\eqref{eq:flux} with a diffusion coefficient as given by Eq.~\eqref{eq:diff} (dashed-dotted lines) normalized by the one across a constant-section channel $\chi_o=D\beta/4L$, as function of the geometry of the channel $\Delta S=\ln\frac{h_0+h_1}{h_0-h_1}=\ln\frac{h_{max}}{h_{min}}$ for different values of the particle radius. Upper right: ratio of $\tilde \chi$ over $\chi$ normalized by $\chi$ for the data sets shown in the left panel.
Bottom: permeability, $\chi$, normalized by the one across a constant-section
channel $\chi_o=D\beta/4L$, as function of the radius of the particle, $R$,
normalized by the average channel width, $h_0$, for different channel geometries
captured by $\Delta S$.  Bottom right: ratio of $\tilde \chi$ over $\chi$
normalized by $\chi$ for the data sets shown in the left panel.
}
\label{fig:chi}
\end{figure*}
\subsection{Piece-wise linear potential and homogeneous diffusion coefficient}
In order to get analytical insight it can be useful to approximate the effective potential $A(x)$ by 
\begin{align}
A_{eq}(x)=-\frac{\Delta A_{eq}}{L}|x|\,,
\label{eq:A_lin}
\end{align}
where
\begin{align}
\Delta A_{eq}=A^{max}_{eq}-A^{min}_{eq}
\end{align}
is the piece-wise linear difference between the maximum and minimum values of $A_{eq}$. 
Moreover, if we assume that the diffusion coefficient is homogeneous
\begin{align}
D(x)=D_0
\end{align}
we get
\begin{align}
\int_{-L}^{L}e^{\beta A_{eq}(x)}dx=&\frac{2L}{\beta \Delta A_{eq}}\left(1-e^{-\beta \Delta A_{eq}}\right)\\
\int_{-L}^{L}e^{-\beta A_{eq}(x)}dx=&\frac{2L}{\beta \Delta A_{eq}}\left(e^{\beta \Delta A_{eq}}-1\right)
\end{align}
%Then we use the normalization condition to determine $\tilde{p}$
%\begin{align}
%\tilde{p}=\frac{\beta \Delta A_{eq}}{2L}%\left(e^{\beta \Delta A_{eq}}-1\right)^{-1}
%\end{align}
and finally by substituting the last expressions into Eq.~\eqref{eq:chi} we obtain an approximated expression for the permeability
\begin{align}
\tilde\chi = \frac{D\beta}{4 L}  \dfrac{\left(\beta\Delta A_{eq}\right)^2}{\cosh(\beta \Delta A_{eq})-1}\,.
\label{eq:flux-simple}
\end{align}
%\JH{$\tilde\chi$ vs. $\chi$?}\JH{Explain Eq46 better} 
Interestingly, Eq.~\eqref{eq:flux-simple} shows that $\chi$ is an even function of $\Delta A_{eq}$. This implies that the transport is insensitive upon flipping the sign of the free energy barrier $\Delta A$. Finally, Eq.\eqref{eq:flux-simple} shows that $\chi$ decays exponentially with $\beta \Delta A_{eq}$.

\section{Discussion}
The reliability of the Fick-Jacobs approximation, namely Eq.\eqref{eq:flux},
has been addressed for point-like particles and it has shown good quantitative
agreement for forces up to $\beta f L\simeq 10$~\cite{Burada2007}. However,
Eq.~\eqref{eq:flux} still requires to numerically compute integrals, whereas
Eq.~\eqref{eq:flux-simple} provides a direct (yet approximated) dependence of $\tilde\chi$ on $\Delta A$. 
Therefore, it is important to address the reliability of
Eq.~\eqref{eq:flux-simple} as compared to the full solution Eq.~\eqref{eq:flux}.
Indeed, all the panels of Fig.~\ref{fig:chi} show that the permeability
calculated with the piece-wise linear model, Eq.~\eqref{eq:flux-simple}, shows
some discrepancies as compared to the full expression give in
Eq.~\eqref{eq:flux}. In particular, as shown in Fig.~\ref{fig:chi} for the case
under consideration ($h_0/L=0.1$) the corrections due to the inhomogeneous
diffusion (dashed-dotted lines) are indistinguishable from those with constant
diffusion coefficient (dashed lies) and hence they do not improve the
approximation. 
On the other hand Fig.~\ref{fig:chi} shows that the simple formula in
Eq.~\eqref{eq:flux-simple} is sufficient  to properly capture the trends and
indeed can be used to estimate the transport of colloidal particle across
porous media.  Interestingly, concerning the magnitude of $\chi$, the bottom
panels of Fig.~\ref{fig:chi} show that the channel permeability decreases upon
increasing the particle size. Interestingly, the decrease is almost linear for
larger corrugations of the channel (larger values of $\Delta S$) whereas for
smaller values of the corrugation it plateaus at smaller values of $R$.

\begin{figure}
\includegraphics[scale=0.5]{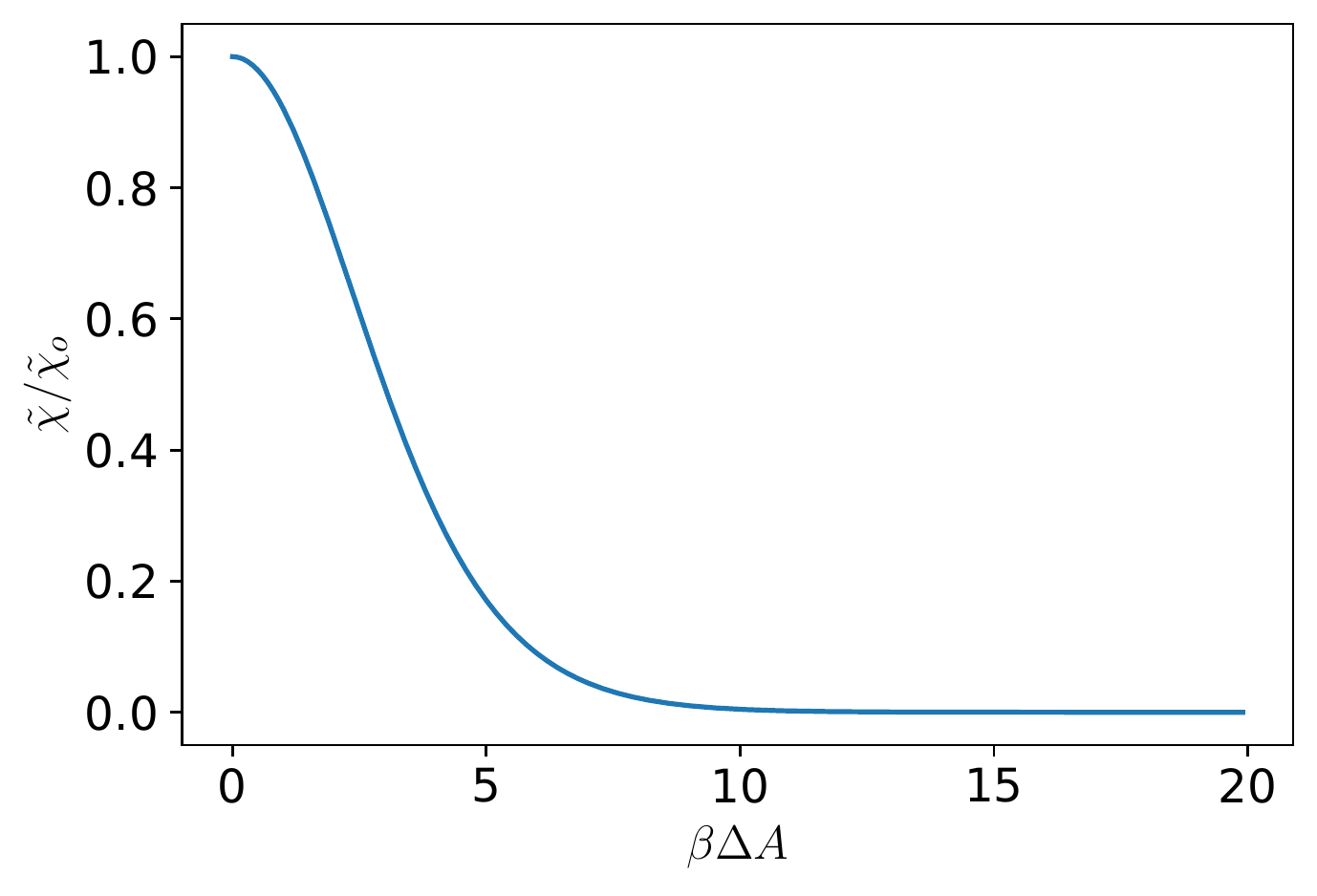}
\caption{Dependence of the approximated channel permeability, $\tilde\chi$, (as defined in Eq.~\eqref{eq:flux-simple}) normalized by that of a constant section channel, $\chi_o$ as function of the amplitude of the dimensionless free energy barrier $\beta \Delta A$ which encodes the physical properties of the confined system.}
\label{fig:chi_DA}
\end{figure}
Finally, we discuss the dependence of $\tilde\chi$ in $\beta \Delta A$ as per
Eq.~\eqref{eq:flux-simple}. As shown in Fig.~\eqref{fig:chi_DA}, $\tilde\chi$ has a
maximum for $\beta \Delta A=0$ and then it decays exponentially for larger
values of $\beta \Delta A$. Interestingly, $\tilde\chi$ attains values close to unity
up to $\beta \Delta A\simeq 5$, i.e. for a free energy barrier much larger than
the thermal energy. 

The fact that Eq.~\eqref{eq:flux-simple} depends solely on $\Delta A$ allows one
to estimate the transport also in situations in which the particles may have
some soft interactions with the walls, like electrostatic interactions. In that
case the free energy barrier will depend not only on the size of the particle
and on the geometry of the channel but also on the charge of both the particle
and the walls of the channels~\cite{Malgaretti_macromolecules,Malgaretti2016}.
Moreover, Eq.~\eqref{eq:flux-simple} allows also to predict the transport of
soft or deformable objects, like proteins or
polymers~\cite{Bianco2016,Malgaretti2019,Carusela2021}.

\section{Conclusions}
We have derived closed formulas for the transport within linear response theory
as well as for higher order corrections.  In particular, we have shown that for
the case of non-interacting systems confined in fore-aft symmetric  channels
the higher order corrections in both the flux and in the density are
identically zero.  Hence, for fore-aft symmetric channels, the full expression
for the flux is indeed the one obtained within the linear response regime.
Accordingly, the channel permeability derived within linear response,
Eq.~\eqref{eq:chi}, is related to the well known expression of the effective diffusion coefficient reported in the literature~\cite{Lifson1962,Rubi2001}.  Moreover, we have shown that, within the linear response, the 
formula for the permeability $\chi$, Eq.~\eqref{eq:chi}, can be further simplified by approximating the local free energy  by piece-wise linear
potential (Eq.~\eqref{eq:A_lin}) to obtain  Eq.~\eqref{eq:flux-simple}, whose overall drop is determined
by the values of the free energy at the bottleneck and at the waist of the
channel. We have shown that such an approximation provides the correct
trends and it is reliable within $\simeq \pm 50\%$ as shown in the right panels of Fig.~\ref{fig:chi}. This feature is crucial since
Eq.~\eqref{eq:flux-simple} can be easily computed and it is valid for all
soft-interactions between the particle and the channel walls.

\section*{Acknowledgments}
We thank I. Pagonabarraga  and  J. Schmid for insightful discussions and acknowledge funding by the Deutsche Forschungsgemeinschaft (DFG, German Research Foundation) – Project-ID 416229255 -- SFB 1411 and Project-ID 431791331 -- SFB 1452. 

\bibliography{simple_biblio,intro_biblio}

\begin{thebibliography}{85}
\expandafter\ifx\csname natexlab\endcsname\relax\def\natexlab#1{#1}\fi
\expandafter\ifx\csname bibnamefont\endcsname\relax
  \def\bibnamefont#1{#1}\fi
\expandafter\ifx\csname bibfnamefont\endcsname\relax
  \def\bibfnamefont#1{#1}\fi
\expandafter\ifx\csname citenamefont\endcsname\relax
  \def\citenamefont#1{#1}\fi
\expandafter\ifx\csname url\endcsname\relax
  \def\url#1{\texttt{#1}}\fi
\expandafter\ifx\csname urlprefix\endcsname\relax\def\urlprefix{URL }\fi
\providecommand{\bibinfo}[2]{#2}
\providecommand{\eprint}[2][]{\url{#2}}

\bibitem[{\citenamefont{Lighthill and Whitham}(1955)}]{Lighthill1955}
\bibinfo{author}{\bibfnamefont{M.~J.} \bibnamefont{Lighthill}}
  \bibnamefont{and} \bibinfo{author}{\bibfnamefont{G.~B.}
  \bibnamefont{Whitham}}, \bibinfo{journal}{Proceedings of the Royal Society of
  London. Series A. Mathematical and Physical Sciences}
  \textbf{\bibinfo{volume}{229}}, \bibinfo{pages}{317} (\bibinfo{year}{1955}).

\bibitem[{\citenamefont{Wang et~al.}(2013)\citenamefont{Wang, Quddus, and
  Ison}}]{Wang_traffic}
\bibinfo{author}{\bibfnamefont{C.}~\bibnamefont{Wang}},
  \bibinfo{author}{\bibfnamefont{M.~A.} \bibnamefont{Quddus}},
  \bibnamefont{and} \bibinfo{author}{\bibfnamefont{S.~G.} \bibnamefont{Ison}},
  \bibinfo{journal}{Safety Science} \textbf{\bibinfo{volume}{57}},
  \bibinfo{pages}{264} (\bibinfo{year}{2013}), ISSN \bibinfo{issn}{0925-7535}.

\bibitem[{\citenamefont{Vermuyten et~al.}(2016)\citenamefont{Vermuyten, Belian,
  {De Boeck}, Reniers, and Wauters}}]{Vermuyten_review}
\bibinfo{author}{\bibfnamefont{H.}~\bibnamefont{Vermuyten}},
  \bibinfo{author}{\bibfnamefont{J.}~\bibnamefont{Belian}},
  \bibinfo{author}{\bibfnamefont{L.}~\bibnamefont{{De Boeck}}},
  \bibinfo{author}{\bibfnamefont{G.}~\bibnamefont{Reniers}}, \bibnamefont{and}
  \bibinfo{author}{\bibfnamefont{T.}~\bibnamefont{Wauters}},
  \bibinfo{journal}{Safety Science} \textbf{\bibinfo{volume}{87}},
  \bibinfo{pages}{167} (\bibinfo{year}{2016}), ISSN \bibinfo{issn}{0925-7535}.

\bibitem[{\citenamefont{Jeong et~al.}(2018)\citenamefont{Jeong, Jun, Cheon, and
  Park}}]{Jeong2018}
\bibinfo{author}{\bibfnamefont{H.~Y.} \bibnamefont{Jeong}},
  \bibinfo{author}{\bibfnamefont{S.-C.} \bibnamefont{Jun}},
  \bibinfo{author}{\bibfnamefont{J.-Y.} \bibnamefont{Cheon}}, \bibnamefont{and}
  \bibinfo{author}{\bibfnamefont{M.}~\bibnamefont{Park}},
  \bibinfo{journal}{Geosciences Journal} \textbf{\bibinfo{volume}{22}},
  \bibinfo{pages}{667} (\bibinfo{year}{2018}).

\bibitem[{\citenamefont{J\"ager et~al.}(2018)\citenamefont{J\"ager, Mendoza,
  and Herrmann}}]{Jaeger2018}
\bibinfo{author}{\bibfnamefont{R.}~\bibnamefont{J\"ager}},
  \bibinfo{author}{\bibfnamefont{M.}~\bibnamefont{Mendoza}}, \bibnamefont{and}
  \bibinfo{author}{\bibfnamefont{H.~J.} \bibnamefont{Herrmann}},
  \bibinfo{journal}{Phys. Rev. Fluids} \textbf{\bibinfo{volume}{3}},
  \bibinfo{pages}{074302} (\bibinfo{year}{2018}).

\bibitem[{\citenamefont{Marin et~al.}(2018)\citenamefont{Marin, Lhuissier,
  Rossi, and K\"ahler}}]{Marin2018}
\bibinfo{author}{\bibfnamefont{A.}~\bibnamefont{Marin}},
  \bibinfo{author}{\bibfnamefont{H.}~\bibnamefont{Lhuissier}},
  \bibinfo{author}{\bibfnamefont{M.}~\bibnamefont{Rossi}}, \bibnamefont{and}
  \bibinfo{author}{\bibfnamefont{C.~J.} \bibnamefont{K\"ahler}},
  \bibinfo{journal}{Phys. Rev. E} \textbf{\bibinfo{volume}{97}},
  \bibinfo{pages}{021102} (\bibinfo{year}{2018}).

\bibitem[{\citenamefont{Kusters et~al.}(2014)\citenamefont{Kusters, van~der
  Heijden, Kaoui, Harting, and Storm}}]{KVKHS14}
\bibinfo{author}{\bibfnamefont{R.}~\bibnamefont{Kusters}},
  \bibinfo{author}{\bibfnamefont{T.}~\bibnamefont{van~der Heijden}},
  \bibinfo{author}{\bibfnamefont{B.}~\bibnamefont{Kaoui}},
  \bibinfo{author}{\bibfnamefont{J.}~\bibnamefont{Harting}}, \bibnamefont{and}
  \bibinfo{author}{\bibfnamefont{C.}~\bibnamefont{Storm}},
  \bibinfo{journal}{Phys. Rev. E} \textbf{\bibinfo{volume}{90}},
  \bibinfo{pages}{033006} (\bibinfo{year}{2014}).

\bibitem[{\citenamefont{Bielinski et~al.}(2021)\citenamefont{Bielinski, Aouane,
  Harting, and Kaoui}}]{BAHK21}
\bibinfo{author}{\bibfnamefont{C.}~\bibnamefont{Bielinski}},
  \bibinfo{author}{\bibfnamefont{O.}~\bibnamefont{Aouane}},
  \bibinfo{author}{\bibfnamefont{J.}~\bibnamefont{Harting}}, \bibnamefont{and}
  \bibinfo{author}{\bibfnamefont{B.}~\bibnamefont{Kaoui}},
  \bibinfo{journal}{Phys. Rev. E} \textbf{\bibinfo{volume}{104}},
  \bibinfo{pages}{065101} (\bibinfo{year}{2021}).

\bibitem[{\citenamefont{Garcimart\'{\i}n
  et~al.}(2015)\citenamefont{Garcimart\'{\i}n, Pastor, Ferrer, Ramos,
  Mart\'{\i}n-G\'omez, and Zuriguel}}]{Zuriguel2015}
\bibinfo{author}{\bibfnamefont{A.}~\bibnamefont{Garcimart\'{\i}n}},
  \bibinfo{author}{\bibfnamefont{J.~M.} \bibnamefont{Pastor}},
  \bibinfo{author}{\bibfnamefont{L.~M.} \bibnamefont{Ferrer}},
  \bibinfo{author}{\bibfnamefont{J.~J.} \bibnamefont{Ramos}},
  \bibinfo{author}{\bibfnamefont{C.}~\bibnamefont{Mart\'{\i}n-G\'omez}},
  \bibnamefont{and} \bibinfo{author}{\bibfnamefont{I.}~\bibnamefont{Zuriguel}},
  \bibinfo{journal}{Phys. Rev. E} \textbf{\bibinfo{volume}{91}},
  \bibinfo{pages}{022808} (\bibinfo{year}{2015}).

\bibitem[{\citenamefont{Altshuler et~al.}(2005)\citenamefont{Altshuler, Ramos,
  Nu\~nez, Fernandez, Batista-Leyva, and Noda}}]{Altshuler2005}
\bibinfo{author}{\bibfnamefont{E.}~\bibnamefont{Altshuler}},
  \bibinfo{author}{\bibfnamefont{O.}~\bibnamefont{Ramos}},
  \bibinfo{author}{\bibfnamefont{Y.}~\bibnamefont{Nu\~nez}},
  \bibinfo{author}{\bibfnamefont{J.}~\bibnamefont{Fernandez}},
  \bibinfo{author}{\bibfnamefont{A.}~\bibnamefont{Batista-Leyva}},
  \bibnamefont{and} \bibinfo{author}{\bibfnamefont{C.}~\bibnamefont{Noda}},
  \bibinfo{journal}{The American Naturalist} \textbf{\bibinfo{volume}{166}},
  \bibinfo{pages}{643} (\bibinfo{year}{2005}).

\bibitem[{\citenamefont{Zuriguel et~al.}(2020)\citenamefont{Zuriguel,
  Echever\'ia, Maza, an\'esar Mart\'in-G\'omez, and
  Garcimar\'in}}]{Zuriguel2020}
\bibinfo{author}{\bibfnamefont{I.}~\bibnamefont{Zuriguel}},
  \bibinfo{author}{\bibfnamefont{I.}~\bibnamefont{Echever\'ia}},
  \bibinfo{author}{\bibfnamefont{D.}~\bibnamefont{Maza}},
  \bibinfo{author}{\bibfnamefont{R.~C.~H.} \bibnamefont{an\'esar
  Mart\'in-G\'omez}}, \bibnamefont{and}
  \bibinfo{author}{\bibfnamefont{A.}~\bibnamefont{Garcimar\'in}},
  \bibinfo{journal}{Safety Science} \textbf{\bibinfo{volume}{121}},
  \bibinfo{pages}{394 } (\bibinfo{year}{2020}), ISSN \bibinfo{issn}{0925-7535},
  \urlprefix\url{http://www.sciencedirect.com/science/article/pii/S0925753519310203}.

\bibitem[{\citenamefont{Squires and Quake}(2005)}]{Squires2005}
\bibinfo{author}{\bibfnamefont{T.~M.} \bibnamefont{Squires}} \bibnamefont{and}
  \bibinfo{author}{\bibfnamefont{S.~R.} \bibnamefont{Quake}},
  \bibinfo{journal}{Rev. Mod. Phys.} \textbf{\bibinfo{volume}{77}},
  \bibinfo{pages}{977} (\bibinfo{year}{2005}).

\bibitem[{\citenamefont{Dressaire and Sauret}(2017)}]{Dressaire2017}
\bibinfo{author}{\bibfnamefont{E.}~\bibnamefont{Dressaire}} \bibnamefont{and}
  \bibinfo{author}{\bibfnamefont{A.}~\bibnamefont{Sauret}},
  \bibinfo{journal}{Soft Matter} \textbf{\bibinfo{volume}{13}},
  \bibinfo{pages}{37} (\bibinfo{year}{2017}).

\bibitem[{\citenamefont{Douf\'ene et~al.}(2019)\citenamefont{Douf\'ene,
  Tourn\'e-P\'eteilh, Etienne, and Aubert-Pou\"essel}}]{Doufene2019}
\bibinfo{author}{\bibfnamefont{K.}~\bibnamefont{Douf\'ene}},
  \bibinfo{author}{\bibfnamefont{C.}~\bibnamefont{Tourn\'e-P\'eteilh}},
  \bibinfo{author}{\bibfnamefont{P.}~\bibnamefont{Etienne}}, \bibnamefont{and}
  \bibinfo{author}{\bibfnamefont{A.}~\bibnamefont{Aubert-Pou\"essel}},
  \bibinfo{journal}{Langmuir} \textbf{\bibinfo{volume}{35}},
  \bibinfo{pages}{12597} (\bibinfo{year}{2019}).

\bibitem[{\citenamefont{Convery and Gadegaard}(2019)}]{Convery2019}
\bibinfo{author}{\bibfnamefont{N.}~\bibnamefont{Convery}} \bibnamefont{and}
  \bibinfo{author}{\bibfnamefont{N.}~\bibnamefont{Gadegaard}},
  \bibinfo{journal}{Micro and Nano Engineering} \textbf{\bibinfo{volume}{2}},
  \bibinfo{pages}{76 } (\bibinfo{year}{2019}), ISSN \bibinfo{issn}{2590-0072},
  \urlprefix\url{http://www.sciencedirect.com/science/article/pii/S2590007219300036}.

\bibitem[{\citenamefont{Weatherall and Willmott}(2015)}]{Willmott2015}
\bibinfo{author}{\bibfnamefont{E.}~\bibnamefont{Weatherall}} \bibnamefont{and}
  \bibinfo{author}{\bibfnamefont{G.~R.} \bibnamefont{Willmott}},
  \bibinfo{journal}{Analyst} \textbf{\bibinfo{volume}{140}},
  \bibinfo{pages}{3318} (\bibinfo{year}{2015}).

\bibitem[{\citenamefont{Saleh and Sohn}(2003)}]{Saleh2003}
\bibinfo{author}{\bibfnamefont{O.~A.} \bibnamefont{Saleh}} \bibnamefont{and}
  \bibinfo{author}{\bibfnamefont{L.~L.} \bibnamefont{Sohn}},
  \bibinfo{journal}{Proc. Natl. Acad. Sci. U. S. A.}
  \textbf{\bibinfo{volume}{100}}, \bibinfo{pages}{820} (\bibinfo{year}{2003}).

\bibitem[{\citenamefont{Ito et~al.}(2004)\citenamefont{Ito, Sun, Bevan, and
  Crooks}}]{Ito2004}
\bibinfo{author}{\bibfnamefont{T.}~\bibnamefont{Ito}},
  \bibinfo{author}{\bibfnamefont{L.}~\bibnamefont{Sun}},
  \bibinfo{author}{\bibfnamefont{M.~A.} \bibnamefont{Bevan}}, \bibnamefont{and}
  \bibinfo{author}{\bibfnamefont{R.~M.} \bibnamefont{Crooks}},
  \bibinfo{journal}{Langmuir} \textbf{\bibinfo{volume}{20}},
  \bibinfo{pages}{6940} (\bibinfo{year}{2004}).

\bibitem[{\citenamefont{Heins et~al.}(2005)\citenamefont{Heins, Siwy, Baker,
  and Martin}}]{Heins2005}
\bibinfo{author}{\bibfnamefont{E.~A.} \bibnamefont{Heins}},
  \bibinfo{author}{\bibfnamefont{Z.~S.} \bibnamefont{Siwy}},
  \bibinfo{author}{\bibfnamefont{L.~A.} \bibnamefont{Baker}}, \bibnamefont{and}
  \bibinfo{author}{\bibfnamefont{R.~C.} \bibnamefont{Martin}},
  \bibinfo{journal}{Nano Lett.} \textbf{\bibinfo{volume}{5}},
  \bibinfo{pages}{1824} (\bibinfo{year}{2005}).

\bibitem[{\citenamefont{Arjmandi et~al.}(2012)\citenamefont{Arjmandi, Van~Roy,
  L., and Borghs}}]{Arjmandi2012}
\bibinfo{author}{\bibfnamefont{N.}~\bibnamefont{Arjmandi}},
  \bibinfo{author}{\bibfnamefont{W.}~\bibnamefont{Van~Roy}},
  \bibinfo{author}{\bibfnamefont{L.}~\bibnamefont{L.}}, \bibnamefont{and}
  \bibinfo{author}{\bibfnamefont{G.}~\bibnamefont{Borghs}},
  \bibinfo{journal}{Anal. Chem.} \textbf{\bibinfo{volume}{84}},
  \bibinfo{pages}{8490} (\bibinfo{year}{2012}).

\bibitem[{\citenamefont{Robards and Ryan}(2022)}]{RobardsBook}
\bibinfo{author}{\bibfnamefont{K.}~\bibnamefont{Robards}} \bibnamefont{and}
  \bibinfo{author}{\bibfnamefont{D.}~\bibnamefont{Ryan}},
  \emph{\bibinfo{title}{Principles and Practice of Modern Chromatographic
  Methods}} (\bibinfo{publisher}{Elsevier}, \bibinfo{address}{Amsterdam},
  \bibinfo{year}{2022}).

\bibitem[{\citenamefont{Reithinger and Arlt}(2011)}]{Reithinger2011}
\bibinfo{author}{\bibfnamefont{M.}~\bibnamefont{Reithinger}} \bibnamefont{and}
  \bibinfo{author}{\bibfnamefont{W.}~\bibnamefont{Arlt}},
  \bibinfo{journal}{Chem. Ing. Tech.} \textbf{\bibinfo{volume}{83}},
  \bibinfo{pages}{83} (\bibinfo{year}{2011}).

\bibitem[{\citenamefont{Michaud et~al.}(2021)\citenamefont{Michaud, Pracht,
  Schilfarth, Damm, Platzer, Haines, Harreiß, Guldi, Spiecker, and
  Peukert}}]{Michaud2021}
\bibinfo{author}{\bibfnamefont{V.}~\bibnamefont{Michaud}},
  \bibinfo{author}{\bibfnamefont{J.}~\bibnamefont{Pracht}},
  \bibinfo{author}{\bibfnamefont{F.}~\bibnamefont{Schilfarth}},
  \bibinfo{author}{\bibfnamefont{C.}~\bibnamefont{Damm}},
  \bibinfo{author}{\bibfnamefont{B.}~\bibnamefont{Platzer}},
  \bibinfo{author}{\bibfnamefont{P.}~\bibnamefont{Haines}},
  \bibinfo{author}{\bibfnamefont{C.}~\bibnamefont{Harreiß}},
  \bibinfo{author}{\bibfnamefont{D.~M.} \bibnamefont{Guldi}},
  \bibinfo{author}{\bibfnamefont{E.}~\bibnamefont{Spiecker}}, \bibnamefont{and}
  \bibinfo{author}{\bibfnamefont{W.}~\bibnamefont{Peukert}},
  \bibinfo{journal}{Nanoscale} \textbf{\bibinfo{volume}{13}},
  \bibinfo{pages}{13116} (\bibinfo{year}{2021}).

\bibitem[{\citenamefont{Seidel-Morgenstern
  et~al.}(2008)\citenamefont{Seidel-Morgenstern, Ke{\ss}ler, and
  Kaspereit}}]{SMKK2008}
\bibinfo{author}{\bibfnamefont{A.}~\bibnamefont{Seidel-Morgenstern}},
  \bibinfo{author}{\bibfnamefont{L.~C.} \bibnamefont{Ke{\ss}ler}},
  \bibnamefont{and}
  \bibinfo{author}{\bibfnamefont{M.}~\bibnamefont{Kaspereit}},
  \bibinfo{journal}{Chemical Engineering \& Technology}
  \textbf{\bibinfo{volume}{31}}, \bibinfo{pages}{826} (\bibinfo{year}{2008}).

\bibitem[{\citenamefont{Soni et~al.}(2010)\citenamefont{Soni, Singer, Yu, Sun,
  McNally, and Meller}}]{Gautam2010}
\bibinfo{author}{\bibfnamefont{G.~V.} \bibnamefont{Soni}},
  \bibinfo{author}{\bibfnamefont{A.}~\bibnamefont{Singer}},
  \bibinfo{author}{\bibfnamefont{Z.}~\bibnamefont{Yu}},
  \bibinfo{author}{\bibfnamefont{Y.}~\bibnamefont{Sun}},
  \bibinfo{author}{\bibfnamefont{B.}~\bibnamefont{McNally}}, \bibnamefont{and}
  \bibinfo{author}{\bibfnamefont{A.}~\bibnamefont{Meller}},
  \bibinfo{journal}{Review of Scientific Instruments}
  \textbf{\bibinfo{volume}{81}}, \bibinfo{pages}{014301}
  (\bibinfo{year}{2010}).

\bibitem[{\citenamefont{Carvalho}(2015)}]{Carvalho2015}
\bibinfo{author}{\bibfnamefont{M.~S.} \bibnamefont{Carvalho}},
  \bibinfo{journal}{Offshore Technology Conference} p.~\bibinfo{pages}{6}
  (\bibinfo{year}{2015}).

\bibitem[{\citenamefont{Foroozesh and Kumar}(2020)}]{Foroozesh2020}
\bibinfo{author}{\bibfnamefont{J.}~\bibnamefont{Foroozesh}} \bibnamefont{and}
  \bibinfo{author}{\bibfnamefont{S.}~\bibnamefont{Kumar}},
  \bibinfo{journal}{Journal of Molecular Liquids}
  \textbf{\bibinfo{volume}{316}}, \bibinfo{pages}{113876}
  (\bibinfo{year}{2020}), ISSN \bibinfo{issn}{0167-7322},
  \urlprefix\url{http://www.sciencedirect.com/science/article/pii/S0167732220324478}.

\bibitem[{\citenamefont{Farhadian and Nikvar-Hassani}(2019)}]{Farhadian2019}
\bibinfo{author}{\bibfnamefont{H.}~\bibnamefont{Farhadian}} \bibnamefont{and}
  \bibinfo{author}{\bibfnamefont{A.}~\bibnamefont{Nikvar-Hassani}},
  \bibinfo{journal}{Bulletin of Engineering Geology and the Environment}
  \textbf{\bibinfo{volume}{78}}, \bibinfo{pages}{3833} (\bibinfo{year}{2019}),
  ISSN \bibinfo{issn}{1435-9537}.

\bibitem[{\citenamefont{Boon and Roij}(2011)}]{Roij2011}
\bibinfo{author}{\bibfnamefont{N.}~\bibnamefont{Boon}} \bibnamefont{and}
  \bibinfo{author}{\bibfnamefont{R.~V.} \bibnamefont{Roij}},
  \bibinfo{journal}{Molecular Physics} \textbf{\bibinfo{volume}{109}},
  \bibinfo{pages}{1229} (\bibinfo{year}{2011}).

\bibitem[{\citenamefont{P.~Preuster}(2017)}]{Preuster2017}
\bibinfo{author}{\bibfnamefont{P.~W.} \bibnamefont{P.~Preuster},
  \bibfnamefont{C.~Papp}}, \bibinfo{journal}{Acc. Chem. Res}
  \textbf{\bibinfo{volume}{50}}, \bibinfo{pages}{74} (\bibinfo{year}{2017}).

\bibitem[{\citenamefont{Solymosi et~al.}(2022)\citenamefont{Solymosi,
  Geißelbrecht, Mayer, Auer, Leicht, Terlinden, Malgaretti, Bösmann,
  Preuster, Harting et~al.}}]{Solimosi2022}
\bibinfo{author}{\bibfnamefont{T.}~\bibnamefont{Solymosi}},
  \bibinfo{author}{\bibfnamefont{M.}~\bibnamefont{Geißelbrecht}},
  \bibinfo{author}{\bibfnamefont{S.}~\bibnamefont{Mayer}},
  \bibinfo{author}{\bibfnamefont{M.}~\bibnamefont{Auer}},
  \bibinfo{author}{\bibfnamefont{P.}~\bibnamefont{Leicht}},
  \bibinfo{author}{\bibfnamefont{M.}~\bibnamefont{Terlinden}},
  \bibinfo{author}{\bibfnamefont{P.}~\bibnamefont{Malgaretti}},
  \bibinfo{author}{\bibfnamefont{A.}~\bibnamefont{Bösmann}},
  \bibinfo{author}{\bibfnamefont{P.}~\bibnamefont{Preuster}},
  \bibinfo{author}{\bibfnamefont{J.}~\bibnamefont{Harting}},
  \bibnamefont{et~al.}, \bibinfo{journal}{Science Advances}
  \textbf{\bibinfo{volume}{8}}, \bibinfo{pages}{eade3262}
  (\bibinfo{year}{2022}).

\bibitem[{\citenamefont{Suter et~al.}(2021)\citenamefont{Suter, Smith, Hack,
  Rasha, Rana, Angel, Shearing, Miller, and Brett}}]{Suter2021}
\bibinfo{author}{\bibfnamefont{T.~A.~M.} \bibnamefont{Suter}},
  \bibinfo{author}{\bibfnamefont{K.}~\bibnamefont{Smith}},
  \bibinfo{author}{\bibfnamefont{J.}~\bibnamefont{Hack}},
  \bibinfo{author}{\bibfnamefont{L.}~\bibnamefont{Rasha}},
  \bibinfo{author}{\bibfnamefont{Z.}~\bibnamefont{Rana}},
  \bibinfo{author}{\bibfnamefont{G.~M.~A.} \bibnamefont{Angel}},
  \bibinfo{author}{\bibfnamefont{P.~R.} \bibnamefont{Shearing}},
  \bibinfo{author}{\bibfnamefont{T.~S.} \bibnamefont{Miller}},
  \bibnamefont{and} \bibinfo{author}{\bibfnamefont{D.~J.~L.}
  \bibnamefont{Brett}}, \bibinfo{journal}{Advanced Energy Materials}
  \textbf{\bibinfo{volume}{11}}, \bibinfo{pages}{2101025}
  (\bibinfo{year}{2021}).

\bibitem[{\citenamefont{Du et~al.}(2022)\citenamefont{Du, Roy, Peach, Turnbull,
  Thiele, and Bock}}]{Du2022}
\bibinfo{author}{\bibfnamefont{N.}~\bibnamefont{Du}},
  \bibinfo{author}{\bibfnamefont{C.}~\bibnamefont{Roy}},
  \bibinfo{author}{\bibfnamefont{R.}~\bibnamefont{Peach}},
  \bibinfo{author}{\bibfnamefont{M.}~\bibnamefont{Turnbull}},
  \bibinfo{author}{\bibfnamefont{S.}~\bibnamefont{Thiele}}, \bibnamefont{and}
  \bibinfo{author}{\bibfnamefont{C.}~\bibnamefont{Bock}},
  \bibinfo{journal}{Chemical Reviews} \textbf{\bibinfo{volume}{122}},
  \bibinfo{pages}{11830} (\bibinfo{year}{2022}).

\bibitem[{\citenamefont{Hepburn et~al.}(2019)\citenamefont{Hepburn, Adlen,
  Beddington, Carter, Fuss, Mac~Dowell, Minx, Smith, and
  Williams}}]{Hepburn2019}
\bibinfo{author}{\bibfnamefont{C.}~\bibnamefont{Hepburn}},
  \bibinfo{author}{\bibfnamefont{E.}~\bibnamefont{Adlen}},
  \bibinfo{author}{\bibfnamefont{J.}~\bibnamefont{Beddington}},
  \bibinfo{author}{\bibfnamefont{E.~A.} \bibnamefont{Carter}},
  \bibinfo{author}{\bibfnamefont{S.}~\bibnamefont{Fuss}},
  \bibinfo{author}{\bibfnamefont{N.}~\bibnamefont{Mac~Dowell}},
  \bibinfo{author}{\bibfnamefont{J.~C.} \bibnamefont{Minx}},
  \bibinfo{author}{\bibfnamefont{P.}~\bibnamefont{Smith}}, \bibnamefont{and}
  \bibinfo{author}{\bibfnamefont{C.~K.} \bibnamefont{Williams}},
  \bibinfo{journal}{Nature} \textbf{\bibinfo{volume}{575}}, \bibinfo{pages}{87}
  (\bibinfo{year}{2019}).

\bibitem[{\citenamefont{Alberts et~al.}(2007)\citenamefont{Alberts, Johnson,
  Lewis, Raff, Roberts, and Walter}}]{Albers_book}
\bibinfo{author}{\bibfnamefont{B.}~\bibnamefont{Alberts}},
  \bibinfo{author}{\bibfnamefont{A.}~\bibnamefont{Johnson}},
  \bibinfo{author}{\bibfnamefont{J.}~\bibnamefont{Lewis}},
  \bibinfo{author}{\bibfnamefont{M.}~\bibnamefont{Raff}},
  \bibinfo{author}{\bibfnamefont{K.}~\bibnamefont{Roberts}}, \bibnamefont{and}
  \bibinfo{author}{\bibfnamefont{P.}~\bibnamefont{Walter}},
  \emph{\bibinfo{title}{Molecular Biology of the Cell}}
  (\bibinfo{publisher}{Garland Science}, \bibinfo{address}{Oxford},
  \bibinfo{year}{2007}).

\bibitem[{\citenamefont{Pethig}(1986)}]{Pethig1986}
\bibinfo{author}{\bibfnamefont{R.}~\bibnamefont{Pethig}}, in
  \emph{\bibinfo{booktitle}{Modern Bioelectrochemistry}}, edited by
  \bibinfo{editor}{\bibfnamefont{F.}~\bibnamefont{Gutmann}} \bibnamefont{and}
  \bibinfo{editor}{\bibfnamefont{H.}~\bibnamefont{Keyzer}}
  (\bibinfo{publisher}{Springer US}, \bibinfo{address}{Boston, MA},
  \bibinfo{year}{1986}), pp. \bibinfo{pages}{199--239}, ISBN
  \bibinfo{isbn}{978-1-4613-2105-7},
  \urlprefix\url{https://doi.org/10.1007/978-1-4613-2105-7_7}.

\bibitem[{\citenamefont{Dubyak}(2004)}]{Dubyak2004}
\bibinfo{author}{\bibfnamefont{G.~R.} \bibnamefont{Dubyak}},
  \bibinfo{journal}{Advances in Physiology Education}
  \textbf{\bibinfo{volume}{28}}, \bibinfo{pages}{143} (\bibinfo{year}{2004}).

\bibitem[{\citenamefont{Calero et~al.}(2011)\citenamefont{Calero, Faraudo, and
  Aguilella-Arzo}}]{Calero}
\bibinfo{author}{\bibfnamefont{C.}~\bibnamefont{Calero}},
  \bibinfo{author}{\bibfnamefont{J.}~\bibnamefont{Faraudo}}, \bibnamefont{and}
  \bibinfo{author}{\bibfnamefont{M.}~\bibnamefont{Aguilella-Arzo}},
  \bibinfo{journal}{Phys. Rev. E} \textbf{\bibinfo{volume}{83}},
  \bibinfo{pages}{021908} (\bibinfo{year}{2011}).

\bibitem[{\citenamefont{Peyser et~al.}(2014)\citenamefont{Peyser, Gillespie,
  Roth, and Nonner}}]{Roth2014}
\bibinfo{author}{\bibfnamefont{A.}~\bibnamefont{Peyser}},
  \bibinfo{author}{\bibfnamefont{D.}~\bibnamefont{Gillespie}},
  \bibinfo{author}{\bibfnamefont{R.}~\bibnamefont{Roth}}, \bibnamefont{and}
  \bibinfo{author}{\bibfnamefont{W.}~\bibnamefont{Nonner}},
  \bibinfo{journal}{Biophysical Journal} \textbf{\bibinfo{volume}{107}},
  \bibinfo{pages}{1841} (\bibinfo{year}{2014}).

\bibitem[{\citenamefont{Lee et~al.}(2017)\citenamefont{Lee, Segets, Süß,
  Peukert, Chen, and Pui}}]{Peukert2017}
\bibinfo{author}{\bibfnamefont{H.}~\bibnamefont{Lee}},
  \bibinfo{author}{\bibfnamefont{D.}~\bibnamefont{Segets}},
  \bibinfo{author}{\bibfnamefont{S.}~\bibnamefont{Süß}},
  \bibinfo{author}{\bibfnamefont{W.}~\bibnamefont{Peukert}},
  \bibinfo{author}{\bibfnamefont{S.-C.} \bibnamefont{Chen}}, \bibnamefont{and}
  \bibinfo{author}{\bibfnamefont{D.~Y.} \bibnamefont{Pui}},
  \bibinfo{journal}{Journal of Membrane Science}
  \textbf{\bibinfo{volume}{524}}, \bibinfo{pages}{682} (\bibinfo{year}{2017}),
  ISSN \bibinfo{issn}{0376-7388}.

\bibitem[{\citenamefont{Melnikov et~al.}(2017)\citenamefont{Melnikov, Hulings,
  and Gracheva}}]{Gracheva2017}
\bibinfo{author}{\bibfnamefont{D.~V.} \bibnamefont{Melnikov}},
  \bibinfo{author}{\bibfnamefont{Z.~K.} \bibnamefont{Hulings}},
  \bibnamefont{and} \bibinfo{author}{\bibfnamefont{M.~E.}
  \bibnamefont{Gracheva}}, \bibinfo{journal}{Physical Review E}
  \textbf{\bibinfo{volume}{95}}, \bibinfo{pages}{063105}
  (\bibinfo{year}{2017}).

\bibitem[{\citenamefont{Bacchin}(2018)}]{Bacchin2018}
\bibinfo{author}{\bibfnamefont{P.}~\bibnamefont{Bacchin}},
  \bibinfo{journal}{Membranes} \textbf{\bibinfo{volume}{8}},
  \bibinfo{pages}{10} (\bibinfo{year}{2018}).

\bibitem[{\citenamefont{Berezhkovskii et~al.}(2019)\citenamefont{Berezhkovskii,
  Dagdug, and Bezrukov}}]{Dagdug2019}
\bibinfo{author}{\bibfnamefont{A.~M.} \bibnamefont{Berezhkovskii}},
  \bibinfo{author}{\bibfnamefont{L.}~\bibnamefont{Dagdug}}, \bibnamefont{and}
  \bibinfo{author}{\bibfnamefont{S.~M.} \bibnamefont{Bezrukov}},
  \bibinfo{journal}{J. Chem. Phys.} \textbf{\bibinfo{volume}{151}},
  \bibinfo{pages}{054113} (\bibinfo{year}{2019}).

\bibitem[{\citenamefont{Nipper and Dixon}(2011)}]{Nipper2011}
\bibinfo{author}{\bibfnamefont{M.}~\bibnamefont{Nipper}} \bibnamefont{and}
  \bibinfo{author}{\bibfnamefont{J.}~\bibnamefont{Dixon}},
  \bibinfo{journal}{Cardiovasc. Eng. Technol.} \textbf{\bibinfo{volume}{2}},
  \bibinfo{pages}{296} (\bibinfo{year}{2011}).

\bibitem[{\citenamefont{Wiig and Swartz}(2012)}]{Wiig2012}
\bibinfo{author}{\bibfnamefont{H.}~\bibnamefont{Wiig}} \bibnamefont{and}
  \bibinfo{author}{\bibfnamefont{M.}~\bibnamefont{Swartz}},
  \bibinfo{journal}{Physiol. Rev.} \textbf{\bibinfo{volume}{92}},
  \bibinfo{pages}{1005} (\bibinfo{year}{2012}).

\bibitem[{\citenamefont{Yoganathan et~al.}(1988)\citenamefont{Yoganathan, Cape,
  Sung, Williams, and Jimoh}}]{Yoganathan1344}
\bibinfo{author}{\bibfnamefont{A.~P.} \bibnamefont{Yoganathan}},
  \bibinfo{author}{\bibfnamefont{E.~G.} \bibnamefont{Cape}},
  \bibinfo{author}{\bibfnamefont{H.-W.} \bibnamefont{Sung}},
  \bibinfo{author}{\bibfnamefont{F.~P.} \bibnamefont{Williams}},
  \bibnamefont{and} \bibinfo{author}{\bibfnamefont{A.}~\bibnamefont{Jimoh}},
  \bibinfo{journal}{Journal of the American College of Cardiology}
  \textbf{\bibinfo{volume}{12}}, \bibinfo{pages}{1344} (\bibinfo{year}{1988}),
  ISSN \bibinfo{issn}{0735-1097},
  \eprint{https://www.onlinejacc.org/content/12/5/1344.full.pdf},
  \urlprefix\url{https://www.onlinejacc.org/content/12/5/1344}.

\bibitem[{\citenamefont{Jensen et~al.}(2016)\citenamefont{Jensen,
  Berg-S{\o}rensen, Bruus, Holbrook, Liesche, Schulz, Zwieniecki, and
  Bohr}}]{RMP_Plants2016}
\bibinfo{author}{\bibfnamefont{K.~H.} \bibnamefont{Jensen}},
  \bibinfo{author}{\bibfnamefont{K.}~\bibnamefont{Berg-S{\o}rensen}},
  \bibinfo{author}{\bibfnamefont{H.}~\bibnamefont{Bruus}},
  \bibinfo{author}{\bibfnamefont{N.~M.} \bibnamefont{Holbrook}},
  \bibinfo{author}{\bibfnamefont{J.}~\bibnamefont{Liesche}},
  \bibinfo{author}{\bibfnamefont{A.}~\bibnamefont{Schulz}},
  \bibinfo{author}{\bibfnamefont{M.~A.} \bibnamefont{Zwieniecki}},
  \bibnamefont{and} \bibinfo{author}{\bibfnamefont{T.}~\bibnamefont{Bohr}},
  \bibinfo{journal}{Rev. Mod. Phys.} \textbf{\bibinfo{volume}{88}},
  \bibinfo{pages}{035007} (\bibinfo{year}{2016}).

\bibitem[{\citenamefont{Shimmen and Yokota}(2004)}]{Shimmen2004}
\bibinfo{author}{\bibfnamefont{T.}~\bibnamefont{Shimmen}} \bibnamefont{and}
  \bibinfo{author}{\bibfnamefont{E.}~\bibnamefont{Yokota}},
  \bibinfo{journal}{Current Opinion in Cell Biology} p.~\bibinfo{pages}{68}
  (\bibinfo{year}{2004}).

\bibitem[{\citenamefont{Zwanzig}(1992)}]{Zwanzig}
\bibinfo{author}{\bibfnamefont{R.}~\bibnamefont{Zwanzig}}, \bibinfo{journal}{J.
  Phys. Chem.} \textbf{\bibinfo{volume}{96}}, \bibinfo{pages}{3926}
  (\bibinfo{year}{1992}).

\bibitem[{\citenamefont{Reguera and Rubi}(2001)}]{Reguera2001}
\bibinfo{author}{\bibfnamefont{D.}~\bibnamefont{Reguera}} \bibnamefont{and}
  \bibinfo{author}{\bibfnamefont{J.~M.} \bibnamefont{Rubi}},
  \bibinfo{journal}{Phys. Rev. E} \textbf{\bibinfo{volume}{64}},
  \bibinfo{pages}{061106} (\bibinfo{year}{2001}).

\bibitem[{\citenamefont{Kalinay and Percus}(2005{\natexlab{a}})}]{Kalinay2005}
\bibinfo{author}{\bibfnamefont{P.}~\bibnamefont{Kalinay}} \bibnamefont{and}
  \bibinfo{author}{\bibfnamefont{J.~K.~P.} \bibnamefont{Percus}},
  \bibinfo{journal}{J. Chem. Phys.} \textbf{\bibinfo{volume}{122}},
  \bibinfo{pages}{204701} (\bibinfo{year}{2005}{\natexlab{a}}).

\bibitem[{\citenamefont{Kalinay and
  Percus}(2005{\natexlab{b}})}]{Kalinay2005_2}
\bibinfo{author}{\bibfnamefont{P.}~\bibnamefont{Kalinay}} \bibnamefont{and}
  \bibinfo{author}{\bibfnamefont{J.~K.} \bibnamefont{Percus}},
  \bibinfo{journal}{Phys. Rev. E} \textbf{\bibinfo{volume}{72}},
  \bibinfo{pages}{061203} (\bibinfo{year}{2005}{\natexlab{b}}).

\bibitem[{\citenamefont{Kalinay and Percus}(2008)}]{Kalinay2008}
\bibinfo{author}{\bibfnamefont{P.}~\bibnamefont{Kalinay}} \bibnamefont{and}
  \bibinfo{author}{\bibfnamefont{J.~K.} \bibnamefont{Percus}},
  \bibinfo{journal}{Phys. Rev. E} \textbf{\bibinfo{volume}{78}},
  \bibinfo{pages}{021103} (\bibinfo{year}{2008}).

\bibitem[{\citenamefont{Martens et~al.}(2011)\citenamefont{Martens, Schmid,
  Schimansky-Geier, and H\"anggi}}]{Martens2011}
\bibinfo{author}{\bibfnamefont{S.}~\bibnamefont{Martens}},
  \bibinfo{author}{\bibfnamefont{G.}~\bibnamefont{Schmid}},
  \bibinfo{author}{\bibfnamefont{L.}~\bibnamefont{Schimansky-Geier}},
  \bibnamefont{and} \bibinfo{author}{\bibfnamefont{P.}~\bibnamefont{H\"anggi}},
  \bibinfo{journal}{Phys. Rev. E} \textbf{\bibinfo{volume}{83}},
  \bibinfo{pages}{051135} (\bibinfo{year}{2011}).

\bibitem[{\citenamefont{Chac\'on-Acosta
  et~al.}(2013)\citenamefont{Chac\'on-Acosta, Pineda, and Dagdug}}]{Dagdug2013}
\bibinfo{author}{\bibfnamefont{G.}~\bibnamefont{Chac\'on-Acosta}},
  \bibinfo{author}{\bibfnamefont{I.}~\bibnamefont{Pineda}}, \bibnamefont{and}
  \bibinfo{author}{\bibfnamefont{L.}~\bibnamefont{Dagdug}},
  \bibinfo{journal}{J. Chem. Phys.} \textbf{\bibinfo{volume}{139}},
  \bibinfo{eid}{214115} (\bibinfo{year}{2013}).

\bibitem[{\citenamefont{Malgaretti et~al.}(2013)\citenamefont{Malgaretti,
  Pagonabarraga, and Rubi}}]{Malgaretti2013}
\bibinfo{author}{\bibfnamefont{P.}~\bibnamefont{Malgaretti}},
  \bibinfo{author}{\bibfnamefont{I.}~\bibnamefont{Pagonabarraga}},
  \bibnamefont{and} \bibinfo{author}{\bibfnamefont{J.}~\bibnamefont{Rubi}},
  \bibinfo{journal}{Frontiers in Physics} \textbf{\bibinfo{volume}{1}},
  \bibinfo{pages}{21} (\bibinfo{year}{2013}).

\bibitem[{\citenamefont{Malgaretti et~al.}(2014)\citenamefont{Malgaretti,
  Pagonabarraga, and Rubi}}]{Malgaretti2014}
\bibinfo{author}{\bibfnamefont{P.}~\bibnamefont{Malgaretti}},
  \bibinfo{author}{\bibfnamefont{I.}~\bibnamefont{Pagonabarraga}},
  \bibnamefont{and} \bibinfo{author}{\bibfnamefont{J.~M.} \bibnamefont{Rubi}},
  \bibinfo{journal}{Phys. Rev. Lett} \textbf{\bibinfo{volume}{113}},
  \bibinfo{pages}{128301} (\bibinfo{year}{2014}).

\bibitem[{\citenamefont{Malgaretti et~al.}(2015)\citenamefont{Malgaretti,
  Pagonabarraga, and Rubi}}]{Malgaretti_macromolecules}
\bibinfo{author}{\bibfnamefont{P.}~\bibnamefont{Malgaretti}},
  \bibinfo{author}{\bibfnamefont{I.}~\bibnamefont{Pagonabarraga}},
  \bibnamefont{and} \bibinfo{author}{\bibfnamefont{J.~M.} \bibnamefont{Rubi}},
  \bibinfo{journal}{Macromol. Symposia} \textbf{\bibinfo{volume}{357}},
  \bibinfo{pages}{178} (\bibinfo{year}{2015}).

\bibitem[{\citenamefont{Malgaretti
  et~al.}(2016{\natexlab{a}})\citenamefont{Malgaretti, Pagonabarraga, and
  Miguel~Rubi}}]{Malgaretti2015}
\bibinfo{author}{\bibfnamefont{P.}~\bibnamefont{Malgaretti}},
  \bibinfo{author}{\bibfnamefont{I.}~\bibnamefont{Pagonabarraga}},
  \bibnamefont{and}
  \bibinfo{author}{\bibfnamefont{J.}~\bibnamefont{Miguel~Rubi}},
  \bibinfo{journal}{J. Chem. Phys.} \textbf{\bibinfo{volume}{144}},
  \bibinfo{pages}{034901} (\bibinfo{year}{2016}{\natexlab{a}}).

\bibitem[{\citenamefont{Chinappi and Malgaretti}(2018)}]{Chinappi2018}
\bibinfo{author}{\bibfnamefont{M.}~\bibnamefont{Chinappi}} \bibnamefont{and}
  \bibinfo{author}{\bibfnamefont{P.}~\bibnamefont{Malgaretti}},
  \bibinfo{journal}{Soft Matter} \textbf{\bibinfo{volume}{14}},
  \bibinfo{pages}{9083} (\bibinfo{year}{2018}).

\bibitem[{\citenamefont{Malgaretti et~al.}(2019)\citenamefont{Malgaretti,
  Janssen, Pagonabarraga, and Rubi}}]{Malgaretti2019_JCP}
\bibinfo{author}{\bibfnamefont{P.}~\bibnamefont{Malgaretti}},
  \bibinfo{author}{\bibfnamefont{M.}~\bibnamefont{Janssen}},
  \bibinfo{author}{\bibfnamefont{I.}~\bibnamefont{Pagonabarraga}},
  \bibnamefont{and} \bibinfo{author}{\bibfnamefont{J.~M.} \bibnamefont{Rubi}},
  \bibinfo{journal}{J. Chem. Phys.} \textbf{\bibinfo{volume}{151}},
  \bibinfo{pages}{084902} (\bibinfo{year}{2019}).

\bibitem[{\citenamefont{Reguera et~al.}(2006)\citenamefont{Reguera, Schmid,
  Burada, Rubi, Reimann, and H\"anggi}}]{Reguera2006}
\bibinfo{author}{\bibfnamefont{D.}~\bibnamefont{Reguera}},
  \bibinfo{author}{\bibfnamefont{G.}~\bibnamefont{Schmid}},
  \bibinfo{author}{\bibfnamefont{P.~S.} \bibnamefont{Burada}},
  \bibinfo{author}{\bibfnamefont{J.~M.} \bibnamefont{Rubi}},
  \bibinfo{author}{\bibfnamefont{P.}~\bibnamefont{Reimann}}, \bibnamefont{and}
  \bibinfo{author}{\bibfnamefont{P.}~\bibnamefont{H\"anggi}},
  \bibinfo{journal}{Phys. Rev. Lett.} \textbf{\bibinfo{volume}{96}},
  \bibinfo{pages}{130603} (\bibinfo{year}{2006}).

\bibitem[{\citenamefont{Reguera et~al.}(2012)\citenamefont{Reguera, Luque,
  Burada, Schmid, Rubi, and H\"anggi}}]{Reguera2012}
\bibinfo{author}{\bibfnamefont{D.}~\bibnamefont{Reguera}},
  \bibinfo{author}{\bibfnamefont{A.}~\bibnamefont{Luque}},
  \bibinfo{author}{\bibfnamefont{P.~S.} \bibnamefont{Burada}},
  \bibinfo{author}{\bibfnamefont{G.}~\bibnamefont{Schmid}},
  \bibinfo{author}{\bibfnamefont{J.~M.} \bibnamefont{Rubi}}, \bibnamefont{and}
  \bibinfo{author}{\bibfnamefont{P.}~\bibnamefont{H\"anggi}},
  \bibinfo{journal}{Phys. Rev. Lett.} \textbf{\bibinfo{volume}{108}},
  \bibinfo{pages}{020604} (\bibinfo{year}{2012}).

\bibitem[{\citenamefont{Marini Bettolo~Marconi
  et~al.}(2015)\citenamefont{Marini Bettolo~Marconi, Malgaretti, and
  Pagonabarraga}}]{Marconi2015}
\bibinfo{author}{\bibfnamefont{U.}~\bibnamefont{Marini Bettolo~Marconi}},
  \bibinfo{author}{\bibfnamefont{P.}~\bibnamefont{Malgaretti}},
  \bibnamefont{and}
  \bibinfo{author}{\bibfnamefont{I.}~\bibnamefont{Pagonabarraga}},
  \bibinfo{journal}{J. Chem. Phys.} \textbf{\bibinfo{volume}{143}},
  \bibinfo{pages}{184501} (\bibinfo{year}{2015}).

\bibitem[{\citenamefont{Malgaretti
  et~al.}(2016{\natexlab{b}})\citenamefont{Malgaretti, Pagonabarraga, and
  Rubi}}]{Malgaretti2016_entropy}
\bibinfo{author}{\bibfnamefont{P.}~\bibnamefont{Malgaretti}},
  \bibinfo{author}{\bibfnamefont{I.}~\bibnamefont{Pagonabarraga}},
  \bibnamefont{and} \bibinfo{author}{\bibfnamefont{J.}~\bibnamefont{Rubi}},
  \bibinfo{journal}{Entropy} \textbf{\bibinfo{volume}{18}},
  \bibinfo{pages}{394} (\bibinfo{year}{2016}{\natexlab{b}}).

\bibitem[{\citenamefont{Puertas et~al.}(2018)\citenamefont{Puertas, Malgaretti,
  and Pagonabarraga}}]{Puertas2018}
\bibinfo{author}{\bibfnamefont{A.}~\bibnamefont{Puertas}},
  \bibinfo{author}{\bibfnamefont{P.}~\bibnamefont{Malgaretti}},
  \bibnamefont{and}
  \bibinfo{author}{\bibfnamefont{I.}~\bibnamefont{Pagonabarraga}},
  \bibinfo{journal}{J. Chem. Phys.} \textbf{\bibinfo{volume}{149}},
  \bibinfo{pages}{174908} (\bibinfo{year}{2018}).

\bibitem[{\citenamefont{Malgaretti and Harting}(2021)}]{Malgaretti2021}
\bibinfo{author}{\bibfnamefont{P.}~\bibnamefont{Malgaretti}} \bibnamefont{and}
  \bibinfo{author}{\bibfnamefont{J.}~\bibnamefont{Harting}},
  \bibinfo{journal}{Soft Matter} \textbf{\bibinfo{volume}{17}},
  \bibinfo{pages}{2062} (\bibinfo{year}{2021}).

\bibitem[{\citenamefont{Bianco and Malgaretti}(2016)}]{Bianco2016}
\bibinfo{author}{\bibfnamefont{V.}~\bibnamefont{Bianco}} \bibnamefont{and}
  \bibinfo{author}{\bibfnamefont{P.}~\bibnamefont{Malgaretti}},
  \bibinfo{journal}{J. Chem. Phys.} \textbf{\bibinfo{volume}{145}},
  \bibinfo{pages}{114904} (\bibinfo{year}{2016}).

\bibitem[{\citenamefont{Malgaretti and Oshanin}(2019)}]{Malgaretti2019}
\bibinfo{author}{\bibfnamefont{P.}~\bibnamefont{Malgaretti}} \bibnamefont{and}
  \bibinfo{author}{\bibfnamefont{G.}~\bibnamefont{Oshanin}},
  \bibinfo{journal}{Polymers} \textbf{\bibinfo{volume}{11}},
  \bibinfo{pages}{251} (\bibinfo{year}{2019}).

\bibitem[{\citenamefont{Bodrenko et~al.}(2019)\citenamefont{Bodrenko, Salis,
  Acosta-Gutierrez, and Ceccarelli}}]{Ceccarelli2019}
\bibinfo{author}{\bibfnamefont{I.~V.} \bibnamefont{Bodrenko}},
  \bibinfo{author}{\bibfnamefont{S.}~\bibnamefont{Salis}},
  \bibinfo{author}{\bibfnamefont{S.}~\bibnamefont{Acosta-Gutierrez}},
  \bibnamefont{and}
  \bibinfo{author}{\bibfnamefont{M.}~\bibnamefont{Ceccarelli}},
  \bibinfo{journal}{J. Chem. Phys.} \textbf{\bibinfo{volume}{150}},
  \bibinfo{pages}{211102} (\bibinfo{year}{2019}).

\bibitem[{\citenamefont{Malgaretti and Stark}(2017)}]{Malgaretti2017}
\bibinfo{author}{\bibfnamefont{P.}~\bibnamefont{Malgaretti}} \bibnamefont{and}
  \bibinfo{author}{\bibfnamefont{H.}~\bibnamefont{Stark}},
  \bibinfo{journal}{The Journal of Chemical Physics}
  \textbf{\bibinfo{volume}{146}}, \bibinfo{pages}{174901}
  (\bibinfo{year}{2017}).

\bibitem[{\citenamefont{Kalinay}(2022)}]{Kalinay2022}
\bibinfo{author}{\bibfnamefont{P.}~\bibnamefont{Kalinay}},
  \bibinfo{journal}{Phys. Rev. E} \textbf{\bibinfo{volume}{106}},
  \bibinfo{pages}{044126} (\bibinfo{year}{2022}).

\bibitem[{\citenamefont{Antunes et~al.}(2022)\citenamefont{Antunes, Malgaretti,
  Harting, and Dietrich}}]{Antunes2022}
\bibinfo{author}{\bibfnamefont{G.~C.} \bibnamefont{Antunes}},
  \bibinfo{author}{\bibfnamefont{P.}~\bibnamefont{Malgaretti}},
  \bibinfo{author}{\bibfnamefont{J.}~\bibnamefont{Harting}}, \bibnamefont{and}
  \bibinfo{author}{\bibfnamefont{S.}~\bibnamefont{Dietrich}},
  \bibinfo{journal}{Phys. Rev. Lett.} \textbf{\bibinfo{volume}{129}},
  \bibinfo{pages}{188003} (\bibinfo{year}{2022}).

\bibitem[{\citenamefont{Ledesma-Durán
  et~al.}(2016)\citenamefont{Ledesma-Durán, Hernández-Hernández, and
  Santamaría-Holek}}]{Santamaria2016}
\bibinfo{author}{\bibfnamefont{A.}~\bibnamefont{Ledesma-Durán}},
  \bibinfo{author}{\bibfnamefont{S.~I.} \bibnamefont{Hernández-Hernández}},
  \bibnamefont{and}
  \bibinfo{author}{\bibfnamefont{I.}~\bibnamefont{Santamaría-Holek}},
  \bibinfo{journal}{The Journal of Physical Chemistry C}
  \textbf{\bibinfo{volume}{120}}, \bibinfo{pages}{7810} (\bibinfo{year}{2016}).

\bibitem[{\citenamefont{Chacón-Acosta
  et~al.}(2020)\citenamefont{Chacón-Acosta, Núñez-López, and
  Pineda}}]{Chacon2020}
\bibinfo{author}{\bibfnamefont{G.}~\bibnamefont{Chacón-Acosta}},
  \bibinfo{author}{\bibfnamefont{M.}~\bibnamefont{Núñez-López}},
  \bibnamefont{and} \bibinfo{author}{\bibfnamefont{I.}~\bibnamefont{Pineda}},
  \bibinfo{journal}{J. Chem. Phys.} \textbf{\bibinfo{volume}{152}},
  \bibinfo{pages}{024101} (\bibinfo{year}{2020}).

\bibitem[{\citenamefont{Malgaretti
  et~al.}(2016{\natexlab{c}})\citenamefont{Malgaretti, Pagonabarraga, and
  Miguel~Rubi}}]{Malgaretti2016}
\bibinfo{author}{\bibfnamefont{P.}~\bibnamefont{Malgaretti}},
  \bibinfo{author}{\bibfnamefont{I.}~\bibnamefont{Pagonabarraga}},
  \bibnamefont{and}
  \bibinfo{author}{\bibfnamefont{J.}~\bibnamefont{Miguel~Rubi}},
  \bibinfo{journal}{J. Chem. Phys.} \textbf{\bibinfo{volume}{144}},
  \bibinfo{pages}{034901} (\bibinfo{year}{2016}{\natexlab{c}}).

\bibitem[{\citenamefont{Berezhkovskii et~al.}(2007)\citenamefont{Berezhkovskii,
  Pustovoit, and Bezrukov}}]{berezhkovskii2007diffusion}
\bibinfo{author}{\bibfnamefont{A.~M.} \bibnamefont{Berezhkovskii}},
  \bibinfo{author}{\bibfnamefont{M.~A.} \bibnamefont{Pustovoit}},
  \bibnamefont{and} \bibinfo{author}{\bibfnamefont{S.~M.}
  \bibnamefont{Bezrukov}}, \bibinfo{journal}{J. Chem. Phys.}
  \textbf{\bibinfo{volume}{126}}, \bibinfo{pages}{134706}
  (\bibinfo{year}{2007}).

\bibitem[{\citenamefont{Burada et~al.}(2007)\citenamefont{Burada, Schmid,
  Reguera, Rubi, and H\"anggi}}]{Burada2007}
\bibinfo{author}{\bibfnamefont{P.~S.} \bibnamefont{Burada}},
  \bibinfo{author}{\bibfnamefont{G.}~\bibnamefont{Schmid}},
  \bibinfo{author}{\bibfnamefont{D.}~\bibnamefont{Reguera}},
  \bibinfo{author}{\bibfnamefont{J.~M.} \bibnamefont{Rubi}}, \bibnamefont{and}
  \bibinfo{author}{\bibfnamefont{P.}~\bibnamefont{H\"anggi}},
  \bibinfo{journal}{Phys. Rev. E} \textbf{\bibinfo{volume}{75}},
  \bibinfo{pages}{051111} (\bibinfo{year}{2007}).

\bibitem[{\citenamefont{Berezhkovskii et~al.}(2015)\citenamefont{Berezhkovskii,
  Dagdug, and Bezrukov}}]{dagdug2015}
\bibinfo{author}{\bibfnamefont{A.~M.} \bibnamefont{Berezhkovskii}},
  \bibinfo{author}{\bibfnamefont{L.}~\bibnamefont{Dagdug}}, \bibnamefont{and}
  \bibinfo{author}{\bibfnamefont{S.~M.} \bibnamefont{Bezrukov}},
  \bibinfo{journal}{J. Chem. Phys.} \textbf{\bibinfo{volume}{143}},
  \bibinfo{eid}{164102} (\bibinfo{year}{2015}).

\bibitem[{\citenamefont{Kalinay and Percus}(2006)}]{Kalinay2006}
\bibinfo{author}{\bibfnamefont{P.}~\bibnamefont{Kalinay}} \bibnamefont{and}
  \bibinfo{author}{\bibfnamefont{J.~K.} \bibnamefont{Percus}},
  \bibinfo{journal}{Phys. Rev. E} \textbf{\bibinfo{volume}{74}},
  \bibinfo{pages}{041203} (\bibinfo{year}{2006}).

\bibitem[{\citenamefont{Pineda et~al.}(2012)\citenamefont{Pineda,
  Alvarez-Ramirez, and Dagdug}}]{Dagdug2012_2}
\bibinfo{author}{\bibfnamefont{I.}~\bibnamefont{Pineda}},
  \bibinfo{author}{\bibfnamefont{J.}~\bibnamefont{Alvarez-Ramirez}},
  \bibnamefont{and} \bibinfo{author}{\bibfnamefont{L.}~\bibnamefont{Dagdug}},
  \bibinfo{journal}{J. Chem. Phys.} \textbf{\bibinfo{volume}{137}},
  \bibinfo{pages}{174103} (\bibinfo{year}{2012}).

\bibitem[{\citenamefont{García-Chung et~al.}(2015)\citenamefont{García-Chung,
  Chacón-Acosta, and Dagdug}}]{Dagdug2015_2}
\bibinfo{author}{\bibfnamefont{A.~A.} \bibnamefont{García-Chung}},
  \bibinfo{author}{\bibfnamefont{G.}~\bibnamefont{Chacón-Acosta}},
  \bibnamefont{and} \bibinfo{author}{\bibfnamefont{L.}~\bibnamefont{Dagdug}},
  \bibinfo{journal}{J. Chem. Phys.} \textbf{\bibinfo{volume}{142}},
  \bibinfo{pages}{064105} (\bibinfo{year}{2015}).

\bibitem[{\citenamefont{Lifson and Jackson}(1962)}]{Lifson1962}
\bibinfo{author}{\bibfnamefont{S.}~\bibnamefont{Lifson}} \bibnamefont{and}
  \bibinfo{author}{\bibfnamefont{J.~L.} \bibnamefont{Jackson}},
  \bibinfo{journal}{J. Chem. Phys.} \textbf{\bibinfo{volume}{36}},
  \bibinfo{pages}{2410} (\bibinfo{year}{1962}).

\bibitem[{\citenamefont{Reimann et~al.}(2001)\citenamefont{Reimann, Van~den
  Broeck, Linke, H\"anggi, Rubi, and P\'erez-Madrid}}]{Rubi2001}
\bibinfo{author}{\bibfnamefont{P.}~\bibnamefont{Reimann}},
  \bibinfo{author}{\bibfnamefont{C.}~\bibnamefont{Van~den Broeck}},
  \bibinfo{author}{\bibfnamefont{H.}~\bibnamefont{Linke}},
  \bibinfo{author}{\bibfnamefont{P.}~\bibnamefont{H\"anggi}},
  \bibinfo{author}{\bibfnamefont{J.~M.} \bibnamefont{Rubi}}, \bibnamefont{and}
  \bibinfo{author}{\bibfnamefont{A.}~\bibnamefont{P\'erez-Madrid}},
  \bibinfo{journal}{Phys. Rev. Lett.} \textbf{\bibinfo{volume}{87}},
  \bibinfo{pages}{010602} (\bibinfo{year}{2001}).

\bibitem[{\citenamefont{Carusela et~al.}(2021)\citenamefont{Carusela,
  Malgaretti, and Rubi}}]{Carusela2021}
\bibinfo{author}{\bibfnamefont{M.~F.} \bibnamefont{Carusela}},
  \bibinfo{author}{\bibfnamefont{P.}~\bibnamefont{Malgaretti}},
  \bibnamefont{and} \bibinfo{author}{\bibfnamefont{J.~M.} \bibnamefont{Rubi}},
  \bibinfo{journal}{Phys. Rev. E} \textbf{\bibinfo{volume}{103}},
  \bibinfo{pages}{062102} (\bibinfo{year}{2021}).

\end{thebibliography}

\end{document}